\documentclass[namedreferences,hyperref,optionalrh,solaromanenum]{spr-sola}

\usepackage{graphicx}                    % For eps figures, newer & more powerfull
\usepackage{amssymb}                    % useful mathematical symbols
\usepackage{color}                       % For color text: \color command
\chardef\us=`\_

\begin{document}

\begin{frontmatter}

\title{Quantifying sunspot group nesting with density-based unsupervised clustering}

%%%%%%%%%%%%%%%%%%%%%%%%%%%%%%%%%%%%%%%%%%%%%%%%%%%
%% Authors Names
%
\author[addressref={aff1},email={nurdankarapinar@gmail.com}]{\inits{N}\fnm{Nurdan}~\snm{Karapınar} \orcid{0000-0001-8301-5897}}
\author[addressref={aff2},corref,email={isik@mps.mpg.de}]{\inits{E}\fnm{Emre}~\snm{Işık} \orcid{0000-0001-6163-0653}}
\author[addressref={aff2},email={}]{\inits{NA}\fnm{Natalie A.}~\snm{Krivova} \orcid{0000-0002-1377-3067}}
\author[addressref={aff3},email={}]{\inits{HV}\fnm{Hakan V.}~\snm{Şenavcı} \orcid{0000-0002-8961-277X}}
\address[id=aff1]{Department of Astronomy and Space Sciences, Graduate School of Natural and Applied Sciences, Ankara University, Ankara, T\"urkiye}
\address[id=aff2]{Max-Planck-Institut für Sonnensystemforschung, Justus-von-Liebig-Weg 3, 37077 Göttingen, Germany}
\address[id=aff3]{Department of Astronomy and Space Sciences, Faculty of Science, Ankara University, Ankara, T\"urkiye}
%%%%%%%%%%%%%%%%%%%%%%%%%%%%%%%%%%%%%%%%%%%%%%%%%%%
%% Runningheads
%
\runningauthor{Karapınar et al.}
\runningtitle{Sunspot group nesting}

%%% Abstract 
\begin{abstract}
Sunspot groups often emerge in spatial--temporal clusters, known as nests or complexes of activity.
Quantifying how frequently such nesting occurs is important for understanding the organisation and recurrence of solar magnetic fields.
We introduce an automated approach based on kernel density estimation and DBSCAN clustering to identify nests in the longitude--time domain and to measure the fraction of sunspot groups that belong to them.
The method combines a smooth representation of emergence patterns with a density-based clustering procedure, validated using synthetic solar-like cycles and corrected for variations in data density.

We apply this method to 151 years of sunspot-group observations from the 
Royal Greenwich Observatory 
Photoheliographic Results (RGO, 1874--1976) and Kislovodsk Mountain 
Astronomical Station (KMAS, 1955--2025)
catalogues.\linebreak
Across all cycles and latitude bands, the mean nesting degree is  $\langle D\rangle = 0.61 \pm 0.12$, 
implying that about 60 percent all sunspot groups emerge within nests.
Nesting is strongest at mid-latitudes (10$^\circ$--20$^\circ$), and results from the two independent datasets agree in the period of overlap.
The nesting degree significantly correlates with the solar activity level, 
with the correlation strengthening when small groups are excluded. 
The characteristic inter-nest spacing contracts from $\sim$200-500~Mm at low 
activity to $\sim$60-100~Mm at solar maximum, approaching typical 
sunspot-group dimensions.
The identified nests range from compact clusters to long-lived, drifting structures, offering new quantitative constraints on the persistence and organisation of solar magnetic activity.
\end{abstract}

%%%%%%%%%%%%%%%%%%%%%%%%%%%%%%%%%%%%%%%%%%%%%%%%%%%
%% Keywords
%
%\keywords{}
\end{frontmatter}
%-------------------------------------------------
%%%%%%%%%%%%%%%%%%%%%%%%%%%%%%%%%%%%%%%%%%%%%%%%%%%
%% Sections
%
\section{Introduction}
\label{sec:intro} 

Sunspots appear in the solar photosphere as visible 
signatures of 
magnetic flux bundles emerging from the solar interior.
These bundles typically form structures with
positive and negative polarities, which outline the
footpoints of rising magnetic flux loops \citep{Lidia15,Weber23}.
The emergence of sunspot groups (SGs) is one of the primary indicators of solar magnetic activity at photospheric levels.

It has long been recognised that sunspot groups do not emerge randomly in longitude.
Instead, they tend to cluster in space and time, forming so-called nests or complexes of activity.
Early statistical work by \citet{bec55} and later analyses of synoptic magnetograms by \citet{gai83} highlighted the recurrent nature of these emergence patterns (\citealt{bec55} called them \emph{Fleckenherde}, or 
`spot herds'). 
These clusters, which may persist for several solar 
rotations, were found to exhibit characteristic 
longitudinal motions that deviate from the local 
(differential) rotation profile 
in both directions and occasionally show diverging and 
converging behaviour over time. 

A systematic and quantitative treatment of nesting was 
initiated by \citet{cas86}, who applied cluster analysis 
to the Greenwich sunspot group data. 
Their analysis revealed 
statistically significant
nests typically spanning less than $30^\circ$ in longitude
and persisting up to 15 solar rotations. 
\citet{bro90} extended this work using
three-dimensional (time, longitude, and latitude) clustering analysis of Greenwich data, 
showing that at least one-third of all sunspot appearances belong to intrinsically 
physical clusters, with characteristic longitudinal and latitudinal spreads of $\sim 2^\circ$  and $\sim 1^\circ$, respectively, and
lifetimes of 1--7 months.
The patterns showed hierarchical organisation that were called `nested nests' with component 
separations below $25^\circ$ longitude. 

\citet{Benevolenskaya05} examined activity nests within 
the broader context of complexes of solar activity, 
linking their formation and evolution to the topology 
of the non-axisymmetrical component of the solar magnetic 
field. Her analysis connected nesting patterns to 
large-scale coronal structures and suggested that 
activity complexes may have roots deep in the 
tachocline \citep[see also][]{Raphaldini23}.

SG nesting was also sometimes connected to the so-called active longitudes, which exhibit
persistent activity patterns across multiple rotations with $180^\circ$ separation between 
hemispheric activity herds and flip-flop cycles of 3.7~years \citep{Berdyugina03, Usoskin05, Usoskin07}.

Despite these efforts, the quantitative characterisation of nesting remains challenging \citep{Pojoga02}.
Earlier analyses often relied on subjective criteria or 
methods that assumed specific cluster shapes or rotation 
profiles.
Moreover, the increasing availability of long-term, 
homogeneous sunspot catalogues motivates the development of 
automated and reproducible techniques to quantify clustering 
behaviour. Modern data-driven
machine learning techniques, including non-parametric density 
estimation and unsupervised pattern-recognition methods, 
offer new opportunities to revisit these questions 
\citep{AsensioRamos23}. 

In this work, we present a systematic approach to identify sunspot-group nests in the longitude--time domain and to quantify the degree of nesting across solar cycles. 
We combine a smooth representation of the emergence density with a density-based clustering approach that detects clusters of arbitrary shape, and we validate the method using synthetic solar-like data.
Applying this approach to 151 years of the Royal Greenwich Observatory 
Photoheliographic Results (RGO, 1874--1976) and Kislovodsk Mountain 
Astronomical Station (KMAS, 1955--2025) observations, we investigate the statistical properties of nesting as a function of latitude and solar cycle, assess their significance, and discuss the implications for the organisation of solar magnetic fields.

\section{Data and methodology}
\label{sec:datamethod}
\subsection{Data}
\label{ssec:data}

We analyse sunspot group (SG) data from two complementary observational 
programs. Our primary dataset comprises the Royal Greenwich Observatory 
Photoheliographic Results (RGO, 1874--1976), providing systematic 
observations spanning Cycles 12--19 \citep{Willis13}. To extend coverage 
through the modern era, we use data from the Kislovodsk Mountain 
Astronomical Station (KMAS, 1955--2025), 
covering Cycles 19--25 \citep{Nagovitsyn07}.

Both datasets provide heliographic coordinates (latitude, longitude), 
observed times, and maximum area measurements for each sunspot group, 
making them suitable for spatial-temporal clustering analysis. The RGO 
catalog is the established reference standard for sunspot observations 
during the photographic era, with well-documented observing protocols 
and systematic daily coverage. KMAS maintains comparable observational 
standards using similar measurement techniques.

The datasets overlap during 1954--1976 (Cycles 19--20), enabling 
direct comparison and cross-calibration. Cross-calibration studies 
show that KMAS individual group areas are systematically smaller 
than RGO measurements by approximately 3\%, with a scaling factor 
$b_{\rm RGO-KMAS} = 1.031 \pm 0.056$ \citep{Mandal20}. For our 
analysis, we apply this correction to bring KMAS areas to the RGO 
scale, ensuring homogeneity across the combined dataset. The overlap 
period also enables validation of our nesting methodology across 
independent observational programs (Section~\ref{ssec:dataset_comparison}).

Together, these datasets provide 151 years of sunspot group observations 
(1874--2025) covering 13 complete solar cycles plus the ongoing Cycle 25, 
with RGO contributing Cycles 12--19 and KMAS extending through Cycles 
19-25.

% This focused selection also benefits from the temporal overlap between RGO and KMAS. 
% Furthermore, investigating the maximum phases of these two cycles allows us to examine the peak nesting tendencies under differing cycle strengths, potentially resolving the ongoing debate on the persistence and physical origin of longitudinal nonuniformities in sunspot emergence \citep{Usoskin05, Berdyugina03}. 

% The feature-weighted density-based clustering algorithm offers an important improvement to existing methods in the literature (why?). 

% A similar DBSCAN-based clustering was carried out on SDO/HMI magnetogram data to distinguish active region clusters \citep{chen2025}. (see if that paper gets accepted before we submit)

\subsection{Methods}
\label{ssec:meth}

In the present work, we analyse sunspot-group statistics in the 
longitude-time plane, focusing on emergence locations and their phase speeds 
in the Carrington reference frame, defined to be co-rotating with about 
$\pm 16^\circ$ latitudes with a sidereal period of about 25.4 days. 

To evaluate SG-nesting quantitatively, we employ Kernel Density Estimation 
(KDE) for visualisation of emergence density patterns 
\citep{Scott79}, combined with the 
DBSCAN (Density Based Spatial Clustering of Applications 
with Noise) clustering algorithm \citep{Ester96} for discrete nest identification. The KDE 
provides intuitive visual guidance, while DBSCAN assigns each sunspot 
group to either an identified nest or classifies it as an isolated emergence.
As the main quantity that quantifies nesting, we define the 
\emph{degree of nesting} as the fraction of SGs in nests, derived from each 
latitude slice's time window, which is a full solar cycle.
It is expressed as 
\begin{equation}
    D_{ij} := \frac{N_{\rm nest}}{N},
    \label{eq:D}
\end{equation}
where $N_{\rm nest}$ is the number of SGs in nests and $N$ the total 
number of SGs in a given time-latitude range, represented by the $i$th 
cycle and $j$th latitude slice.

\subsubsection{Kernel Density Estimation}
\label{sssec:kde}

Kernel Density Estimation (KDE) is a non-parametric technique 
for estimating the probability density function of a dataset by 
placing a selected kernel function $K_H$ at each data point and 
summing their contributions \citep{Scott79}. 
For a dataset of $n$ SGs 
with emergence longitudes and times $(\phi_i, t_i)$ and maximum 
observed areas $A_i$, we define area-weighted weights as 
$w_{A,i} = \ln(A_i)/\ln(A_{\rm max})$, where 
$A_{\rm max} = \max_j(A_j)$ is the largest SG area in the analysis 
window. The two-dimensional 
area-weighted kernel density estimator in the longitude-time space is
\begin{equation}
\hat{f}(\phi, t) = \frac{1}{\sum_{i=1}^n w_{A,i}} \sum_{i=1}^n w_{A,i} \, K_H\left(\frac{\phi - \phi_i}{h_\phi}, \frac{t - t_i}{h_t}\right)
\end{equation}
where $K_H$ is a bivariate Gaussian kernel with bandwidth parameters 
$h_\phi$ and $h_t$ determined using Scott's rule \citep{Scott79}. 
Prior to KDE, both 
longitude and time coordinates are standardised using z-score normalisation. 
The area weighting ensures that larger SGs contribute more 
significantly to the density estimation, reflecting their enhanced 
magnetic flux content.

In our analysis, KDE serves primarily as a \emph{visualisation tool} 
to display the underlying density structure of sunspot emergence patterns. 
The discrete identification of nests is performed by the DBSCAN algorithm 
described below.

\subsubsection{Clustering with DBSCAN}
\label{sssec:dbscan}

To identify discrete nesting patterns in SG emergence, 
we employ the Density Based Spatial Clustering of Applications 
with Noise (DBSCAN) algorithm \citep{Ester96}. 
Unlike traditional clustering methods such as k-means, 
DBSCAN does not require specifying the number of clusters beforehand 
and can identify clusters of arbitrary shapes, making it well-suited 
for detecting the elongated or irregular nesting patterns expected in 
solar active region emergence.

DBSCAN operates using two hyperparameters: the neighbourhood radius 
$\varepsilon$ and the minimum number of points $m_p$ required to form 
a dense region. The algorithm classifies data points (each point being 
a sunspot group) into \emph{core points} 
(forming cluster interiors), \emph{border points} 
(on cluster peripheries), and \emph{noise points} (isolated outliers). 
In our context, noise points 
correspond to SGs emerging in isolation, while clustered 
points represent nest members.
% RESULTS: the flux or area distributions of isolated vs. nested SGs.

\subsubsection{Parameter Selection and Validation}
\label{sssec:params}

%The choice of DBSCAN parameters critically affects the identified  clustering structure. 
The DBSCAN parameters $\varepsilon$ (neighbourhood radius) and $m_p$ (minimum number of points) determine the clustering sensitivity and must be chosen to reflect the intrinsic spatial–temporal structure of sunspot-group emergence.
We explored two complementary approaches: $(i)$ an adaptive 
selection informed by the KDE bandwidth, and $(ii)$ a fixed-parameter approach calibrated using synthetic data.
The KDE-informed approach assumes that the smoothing scales appropriate for visualising the emergence density should also approximate the characteristic scales of potential clusters.
%and validated through  injection-recovery tests on synthetic data.
However, validation tests showed that KDE bandwidths respond strongly to the overall spatial–temporal extent of the data rather than to the intrinsic cohesive scales of nests.
As a result, this approach systematically overestimates the clustering strength when the true nesting degree is low ($D \lesssim 0.5$).

\iffalse
The KDE-informed approach uses the bandwidth from density estimation 
to set $\varepsilon$, with the rationale that the smoothing scale 
appropriate for density visualisation should correspond to the 
neighbourhood scale for clustering. However, our validation tests on 
synthetic data with known nesting degrees revealed a systematic bias: 
because KDE bandwidth responds to overall data extent rather than 
intrinsic clustering structure, this approach overestimates nesting 
at low true nesting degrees ($D \lesssim 0.5$) while performing 
adequately at high nesting degrees.
\fi

We therefore adopted a fixed pair of parameters, $\varepsilon = 0.11$ and $m_p = 3$, chosen after 
%in standardised longitude-time space, validated through both the 
%KDE-informed approach mentioned above and 
extensive 
injection-recovery testing (Section~\ref{ssec:validation}).
These values were selected to minimise the mean absolute bias between injected and recovered nesting degrees across the solar-relevant range $0.4\lesssim D\lesssim 0.7$. The choice $m_p = 3$ reflects the minimum plausible number of independently observed sunspot groups required to constitute a physically meaningful nest, consistent with earlier definitions based on longitudinal clustering in active longitudes.

Although a fixed $\varepsilon$ performs well for typical data densities, sparse data windows, particularly high-latitude bands during low-activity periods, can lead DBSCAN to artificially fragment genuine nests.
To account 
for this,
varying data density across different temporal windows and latitude 
bands, 
we introduced a sparsity correction that adjusts the effective radius 
$\varepsilon$: 
\begin{eqnarray}
\varepsilon_{\mathrm{eff}} = 
\left\{
\begin{array}{ll}
\varepsilon_0 \times \min\left(1.5, \sqrt{N_{\mathrm{ref}}/N}\right), & N < N_{\mathrm{ref}} \\
\varepsilon_0, & N \geq N_{\mathrm{ref}}
\end{array}
\right.
% \varepsilon_{\rm eff} = \varepsilon_0 \times \min\left(1.5, \sqrt{\frac{N_{\rm ref}}{N}}\right),
\label{eq:sparsity}
\end{eqnarray}
where $\varepsilon_0 = 0.11$ is the baseline parameter, $N$ is the 
number of SGs in the analysis window, and $N_{\rm ref} = 350$ 
is a reference sample size. 
%MAYBE ONE OF THE BELOW FITS? OT REMOVE THE COMMENT ABOVE
The reference value was determined empirically: synthetic tests showed that windows with $N\gtrsim 350$ yield stable recovery of nests without excessive fragmentation, so the correction compensates only in sparser cases.
%The reference value was determined empirically as the minimum sample size at which injection–recovery tests produced stable clustering without artificial fragmentation.
This correction increases $\varepsilon$ 
by up to 50\% for sparse datasets, preventing excessive fragmentation 
of genuine nesting structures during low-activity periods.
For dense 
data slices with $N > N_{\rm ref}$, no correction is applied 
($\varepsilon_{\rm eff} = \varepsilon_0$).

This hybrid procedure~-- fixed $\varepsilon$ with a bounded sparsity correction~-- produces robust and nearly unbiased recovery of the nesting degree in both synthetic and observational data.

\subsubsection{Analysis Windows and Boundary Handling}

We process SG emergence data in non-overlapping temporal windows 
corresponding to full solar cycles (minimum to minimum), subdivided 
into $5^\circ$ latitude bands from $-35^\circ$ to $+35^\circ$. 
This latitude slicing is motivated by the following observations: 
$(i)$ SG nesting occurs in both longitude and latitude. $(ii)$ 
SG nests can exhibit various rotation rates in the Carrington frame 
and in many cases intrinsic phase motion, but no systematic motion 
in latitude was reported. 
% accounts for the systematic latitude migration of sunspot emergence 
% throughout the solar cycle and ensures comparison of emergence patterns 
% under approximately uniform physical conditions.

A critical consideration is the proper handling of the longitude 
discontinuity at $0^\circ/360^\circ$. 
%SG nests can randomly span this coordinate boundary, so 
Since DBSCAN identifies spatially coherent structures, nests that traverse the discontinuity would be artificially split if the domain were treated naively. To avoid this,
we implement a wraparound correction by 
replicating sunspot groups near the boundary: groups with longitude $\phi > 275^\circ$ are 
duplicated at $\phi- 360^\circ$,
%negative longitudes in the range $(-85^\circ,0^\circ)$, 
and groups with longitude $\phi <85^\circ$ are duplicated beyond $\phi+360^\circ$. 
The clustering is then performed on the propagated dataset, but only groups within the original $[0^\circ, 360^\circ)$ 
range are retained when computing the nesting degree.
This ensures that nests are identified consistently irrespective of their location relative to the coordinate singularity.

\subsubsection{Statistical Significance}

The primary output of our analysis is the nesting degree, 
defined by Eq.~(\ref{eq:D}). 
% \begin{equation}
% D = \frac{N_{\rm nest}}{N_{\rm total}}
% \end{equation}
% where $N_{\rm nest}$ is the number of SGs assigned to 
% identified clusters and $N_{\rm total}$ is the total number of groups 
% in the analysis window.
To validate the statistical significance of identified nests and 
avoid spurious clustering from random alignments in sparse 
data, we test each identified nest against the null hypothesis of uniform 
random emergence. Under this hypothesis, sunspot groups are distributed 
uniformly across the longitude-time domain, and the number of groups 
falling within any subregion follows a Poisson distribution. 
For each cluster occupying a bounding box of area 
$A_c = \Delta\phi \times \Delta t$, 
the expected number of groups under uniform distribution is
\begin{equation}
N_{\rm expected} = \frac{N_{\rm total}}{360^\circ T_{\rm window}} A_c,
\end{equation}
where $N_{\rm total}$ is the total number of SGs in the analysis window 
and $T_{\rm window}$ is its temporal extent. We then calculate the 
probability of observing $n_c$ or more groups within this area by 
chance yields 
\begin{equation}
    p(X \geq n_c | N_{\rm expected}) = 1 - F_{\rm Poisson}(n_c - 1;N_{\rm expected}),
\end{equation}
where $F_{\rm Poisson}$ is the Poisson cumulative distribution function. 
Clusters with $p<0.05$ are classified as statistically significant, 
indicating that their overdensity is unlikely to arise from random 
fluctuations. 

% For each identified 
% cluster containing $n_c$ SGs within a bounding box of area 
% $A_c = \Delta\phi \times \Delta t$, we calculate the expected number 
% under uniform random emergence:
% \begin{equation}
% \lambda = \rho_{\rm window} \times A_c = \frac{N_{\rm total}}{360^\circ \times T_{\rm window}} \times A_c
% \end{equation}
% where $T_{\rm window}$ is the temporal extent of the analysis window. 
% Clusters with Poisson probability $P(X \geq n_c | \lambda) < 0.05$ 
% are classified as statistically significant. We report both the total 
% nesting degree $D$ and the significant nesting degree $D_{\rm sig}$, 
% which counts only groups belonging to significant nests.

\subsection{Validation with Synthetic Data}
\label{ssec:validation}

To validate our methodology and characterise any systematic biases, 
we performed injection-recovery tests using synthetic sunspot data 
with prescribed nesting degrees. The synthetic data generator produces 
realistic SG distributions including: $(i)$ latitude drift 
following the butterfly diagram pattern, $(ii)$ area distributions 
matching observed statistics, and $(iii)$ controlled nesting with 
specified probability $D_{\rm inj}$ that each new group emerges 
within an existing nest rather than at a random location.

\begin{figure}
\centering
\includegraphics[width=\textwidth]{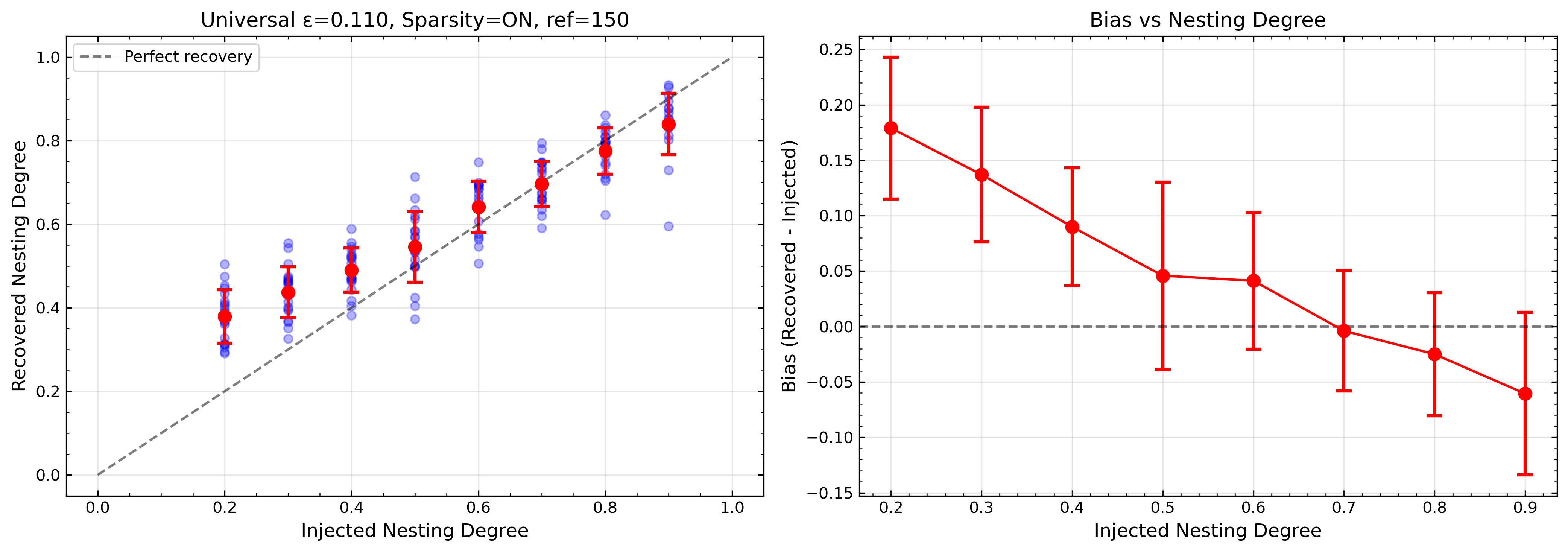}
\caption{Injection-recovery validation of DBSCAN clustering for synthetic 
SG-emergence data of solar-like cycles, showing 
recovered vs. injected nesting degree (left) and the corresponding bias 
as a function of the injected degree (right). 
Red symbols show mean values with error bars indicating standard 
deviation over 160 independent realisations. 
% The fixed-parameter method ($\varepsilon = 0.11$ with sparsity correction) 
% shows good agreement across the range for $D>0.5$, with a 
% mean bias of 0.03 and scatter of 0.02.
}
\label{fig:validation}
\end{figure}

Figure~\ref{fig:validation} shows the results of injection-recovery 
tests for nesting degrees $D_{\rm inj}$ ranging from $0.2$ to $0.9$. 
For each injected value, we generated 20 independent realisations 
and applied DBSCAN with $\varepsilon = 0.11$ and 
sparsity correction (Equation~\ref{eq:sparsity}).

The recovered nesting degree shows good agreement with injected values 
across the tested range, with mean bias  
$\langle D_{\rm rec} - D_{\rm inj} \rangle = $ 
0.062 and 
RMS scatter of 0.10. At low nesting degrees ($D_{\rm inj} < 0.4$),  
slight underestimation occurs. The overall scatter corresponds to 16% 
relative error for typical $D\sim 0.6$. 
In the solar-relevant range 
$D=0.5-0.7$, the scatter reduces to $\sigma = 0.07$ (11\% for $D\sim 0.6$), 
indicating good precision for typical active region nesting degrees.
This validation 
demonstrates that our fixed-parameter approach with sparsity correction 
provides robust nesting degree estimates across the range of values 
expected in solar data.

\iffalse
\subsection{Cross-Validation Between RGO and KMAS}
\label{sec:rgo-kmas}

The RGO and KMAS catalogues overlap during 1954--1976 (Cycles 19--20), enabling a consistency check of our methodology. Although the two datasets differ in instrumentation and measurement procedures, earlier studies have shown that their sunspot-group areas can be cross-calibrated to within a few percent \citep{Mandal20}. After applying the area correction of \citet{Mandal20}, we analysed both datasets separately over the overlap period.

% The resulting nesting degrees agree within their $1-\sigma$ uncertainties for all latitude bands, and no systematic offset is detected. Furthermore, the latitude dependence of the nesting degree (higher at mid-latitudes and lower near the cycle end) is reproduced in both datasets independently. This agreement confirms that the combined RGO+KMAS dataset is suitable for long-term analysis of sunspot-group nesting, and that the method is insensitive to small differences in catalog construction.
\fi

\section{Results}
\label{sec:results}

We applied the validated DBSCAN method described in 
Section~\ref{ssec:meth} to RGO and KMAS SG datasets 
spanning 1878--2025, covering 14 solar cycles (Cycles 12--25). 
The analysis was performed using full solar cycle temporal windows, 
with each cycle analysed separately over $5^\circ$ latitude bands 
from $-35^\circ$ to $+35^\circ$. 
% This yields 98 independent analysis 
% windows (7 cycles with 14 latitude bands).
%This sounds as if you ONLY applied the method to KMAS. I do not think this is the case. I suppose you simply wanted to start with this dataset. (And yes - you do use both later). Thus, please remove thus intro here but move this info into 3.1. You start 3.1. with something like: We first show some examples of the performance of our method on the KMAS dataset. The analysis was performed on.. copy the rest from here.

\subsection{Identification of Sunspot Group Nests}
\label{ssec:nest_identification}

\begin{figure}
    \centering
    \includegraphics[width=0.49\linewidth]{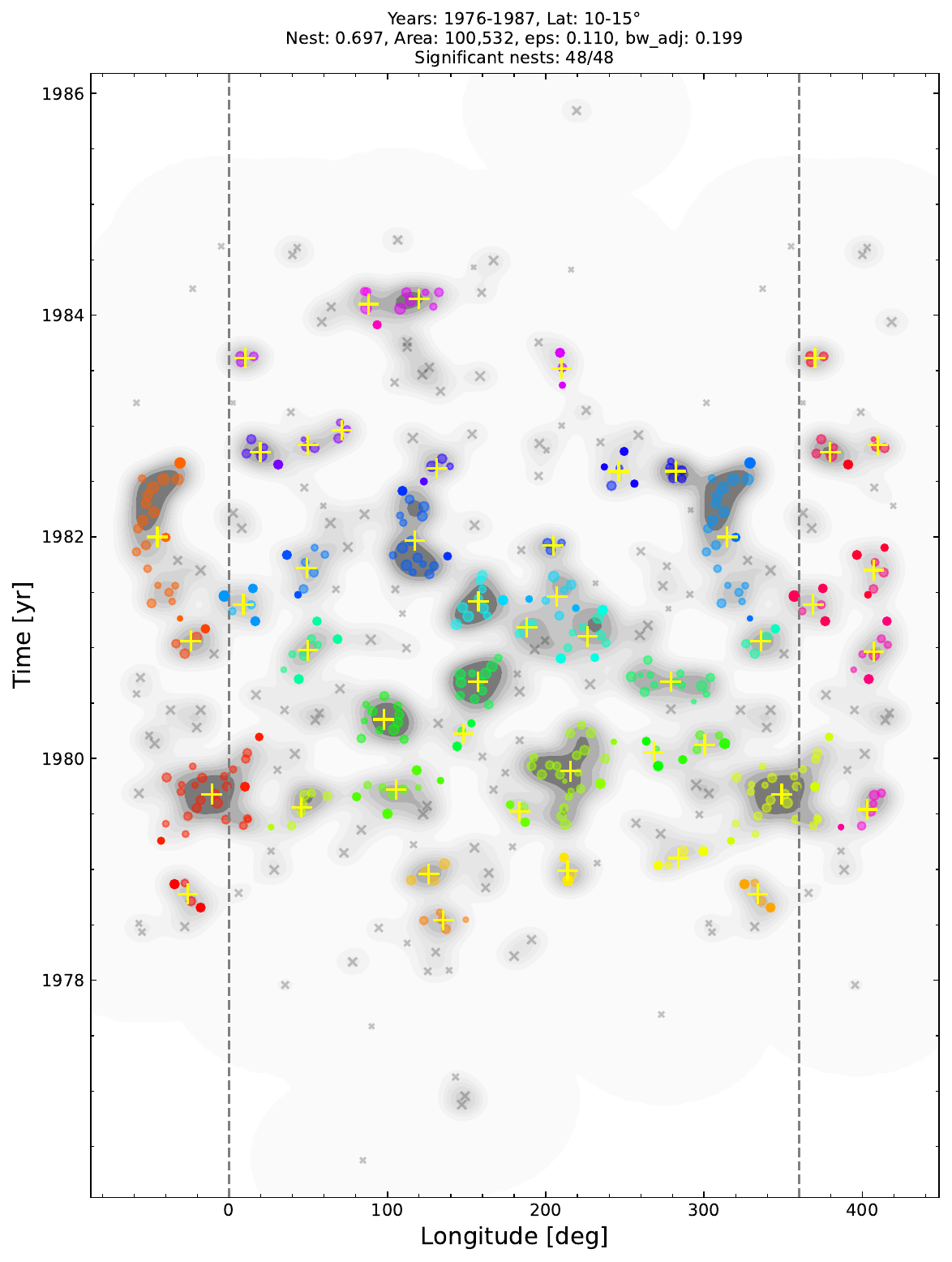}
    \includegraphics[width=0.49\linewidth]{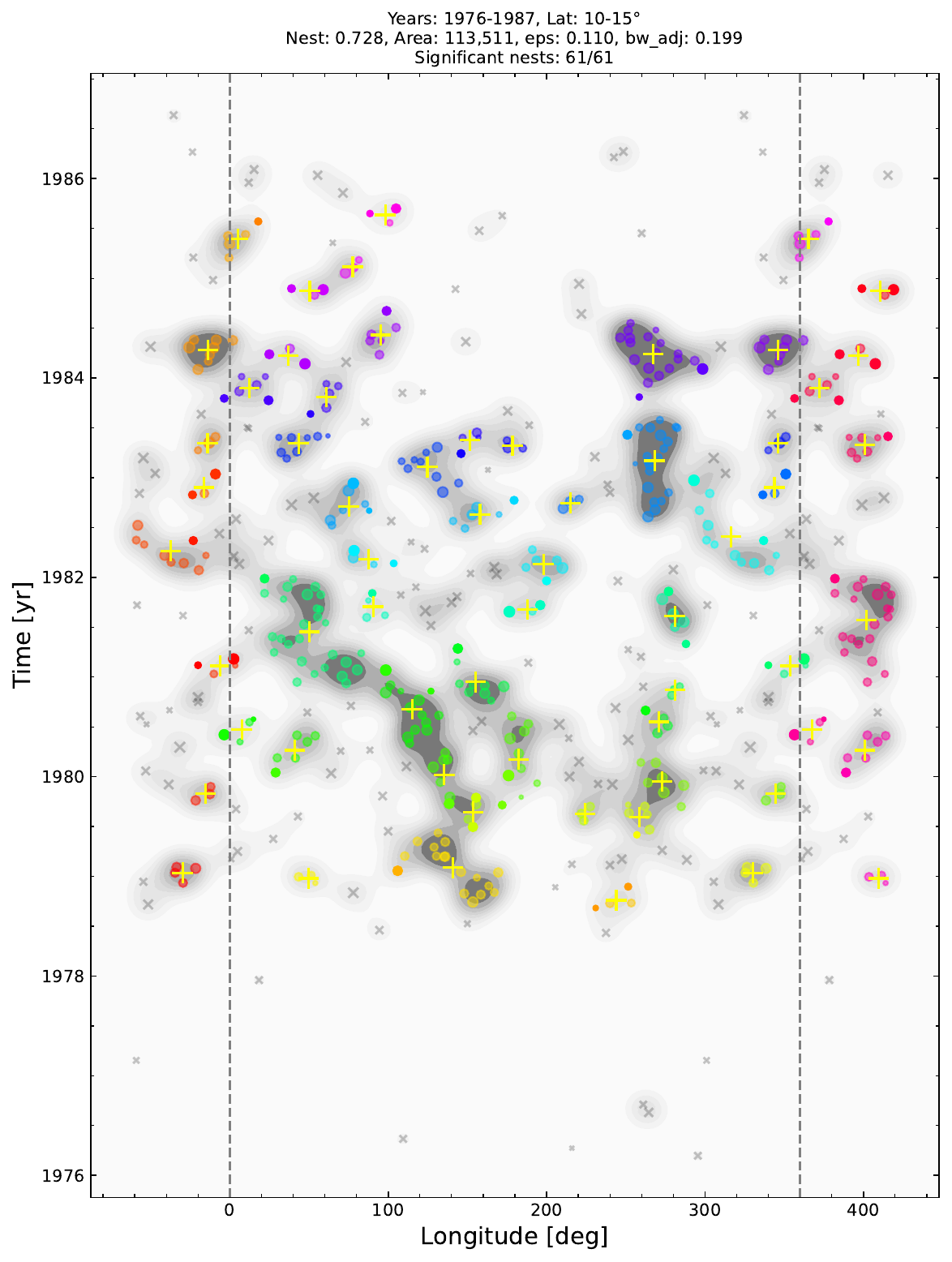}
    \caption{Carrington longitudes and times of SG emergence from KMAS data, during solar 
    cycle 21 across the latitude band $10^\circ-15^\circ$ of the northern 
    (left panel) and southern (right panel) hemisphere. Grey-scale 
    background show the SG-area-weighted KDE density distribution where 
    the areas were taken in their peak value. The coloured circles denote 
    nest members and grey crosses non-members according to the area-weighted 
    DBSCAN. Yellow plus signs show the area-weighted centres of nests. 
    The nesting degree, total SG area, adopted $\varepsilon$, and the 
    adjusted KDE bandwidth (bw\_adjust), and the number of significant vs. 
    total nests are given on plot titles.}
    \label{fig:nest_examples}
\end{figure}

Figure~\ref{fig:nest_examples} presents representative examples of 
identified SG nests on the Carrington longitude-time plane. 
The background contours show the area-weighted KDE density estimation, 
providing visual context for the underlying emergence patterns. 
Coloured points represent SGs assigned to individual nests 
by the DBSCAN algorithm, with cluster centres marked by yellow crosses. 
Black crosses indicate isolated groups not participating in nesting events.

These examples demonstrate the capability of our method to detect 
clustering patterns of varying spatial and temporal extent. Some nests 
are tightly confined (compact clusters), while others show extended 
temporal persistence spanning multiple years, reflecting the long-lived 
nature of subsurface magnetic structures that give rise to repeated 
emergence in preferred longitudes. These preferred longitudes often 
drift in time, shifting the nest centre at various rates and directions, 
% independent of the local (differential) rotation rate. The longitude-time diagrams of the entire 
% period can be reached in the online version (Appendix, available online?). 

\subsection{Cross-Validation Between Datasets}
\label{ssec:dataset_comparison}

To assess the robustness of our methodology and quantify systematic 
differences between observational programmes, we applied our nesting 
analysis independently to both RGO and KMAS data during their overlap 
period (1954--1976, covering Cycles 19--20). Both datasets were processed 
using identical analysis windows and latitude bands ($5^\circ$ intervals), 
with KMAS areas scaled to the RGO reference frame using $b_{\rm RGO-KMAS} = 1.031$ 
\citep{Mandal20}.

\begin{figure}
\centering
\includegraphics[width=0.55\textwidth]{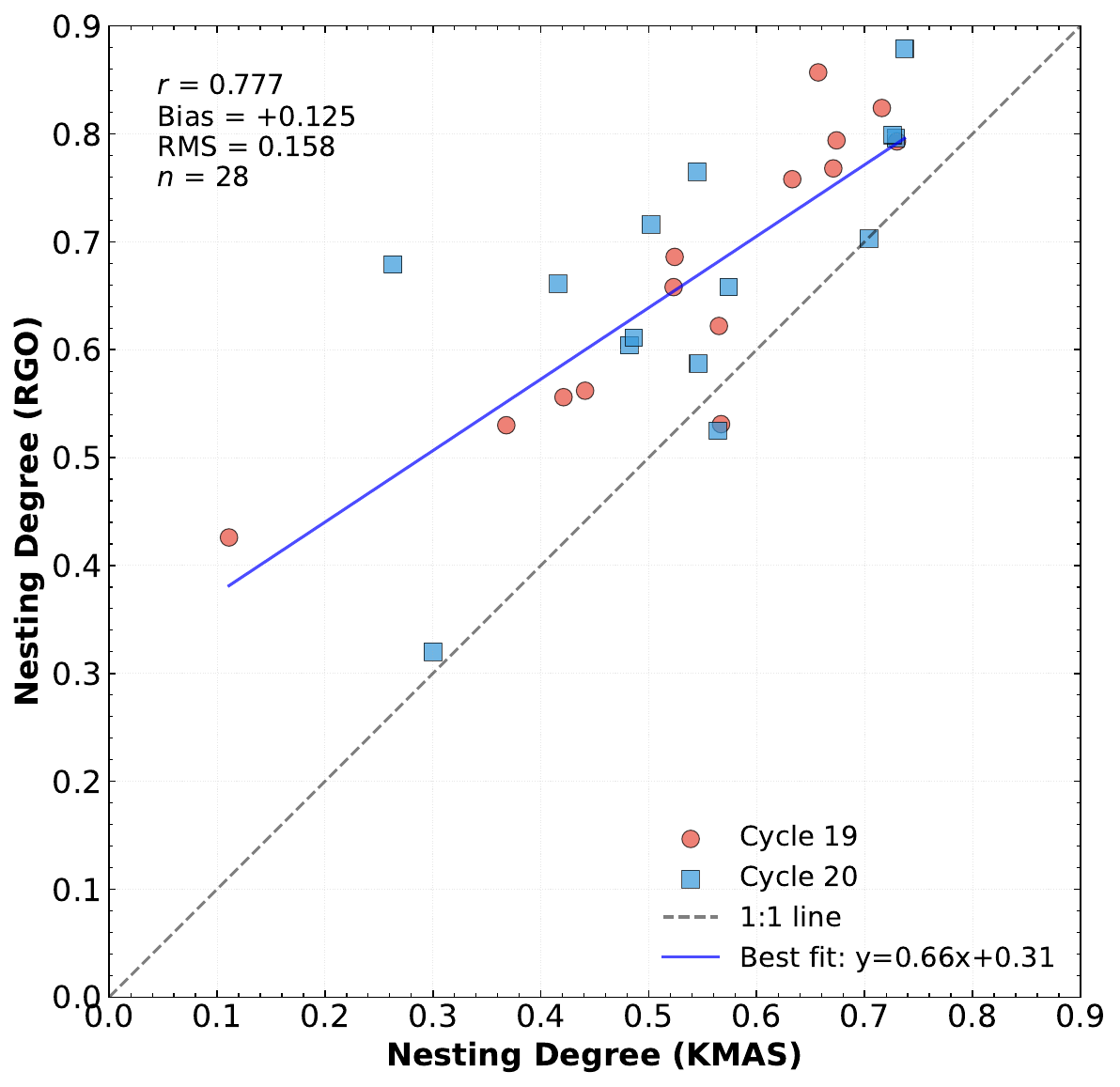}
\caption{Comparison of nesting degrees from RGO and 
KMAS during overlap period (Cycles 19--20, 1954--1976). Red circles: 
Cycle 19 ($n=14$); blue squares: Cycle 20 ($n=14$). Strong correlation 
($r=0.777$) with systematic offset following 
$D_{\rm RGO} = 0.66 \, D_{\rm KMAS} + 0.31$ (RMS = 0.158).}
\label{fig:dataset_comparison}
\end{figure}

Figure~\ref{fig:dataset_comparison} shows the 
comparison of recovered nesting degrees across 28 matched time-latitude windows. 
Despite the area calibration, RGO systematically recovers higher nesting 
degrees. The datasets are strongly correlated ($r = 0.777$, $p < 10^{-6}$), 
demonstrating that both capture the same underlying nesting patterns. 
The relationship follows:
\begin{equation}
D_{\rm RGO} = 0.66 \, D_{\rm KMAS} + 0.31
\label{eq:rgo_kmas_calibration}
\end{equation}
with RMS scatter of 0.158. 
The positive intercept implies that RGO systematically yields higher nesting degrees than KMAS over most of the observed range. The slope below unity indicates that the magnitude of this offset decreases with increasing nesting degree ($\sim 0.28$ at $D_{\rm KMAS} = 0.1$ and $\sim 0.07$ at $D_{\rm KMAS} = 0.7$).
This 
pattern is consistent for both cycles 
(Cycle 19: $r = 0.891$; Cycle 20: $r = 0.657$) and both 
hemispheres (N: mean bias $+0.140$; S: mean bias $+0.106$).

The weaker correlation in Cycle~20 (r = 0.657) 
compared to Cycle~19 (r = 0.891) suggests that systematic 
differences between the datasets might depend on the 
average activity level. 
During 
lower-activity periods (Cycle 20 as compared to 19), RGO's 
superior small-group detection may become more important for 
identifying weak or dispersed nests. 
Alternatively, this difference may partly reflect changes in KMAS observing procedures 
during its early operational phase 
following its start in 1955.

\begin{figure}
\centering
\includegraphics[width=0.49\textwidth]{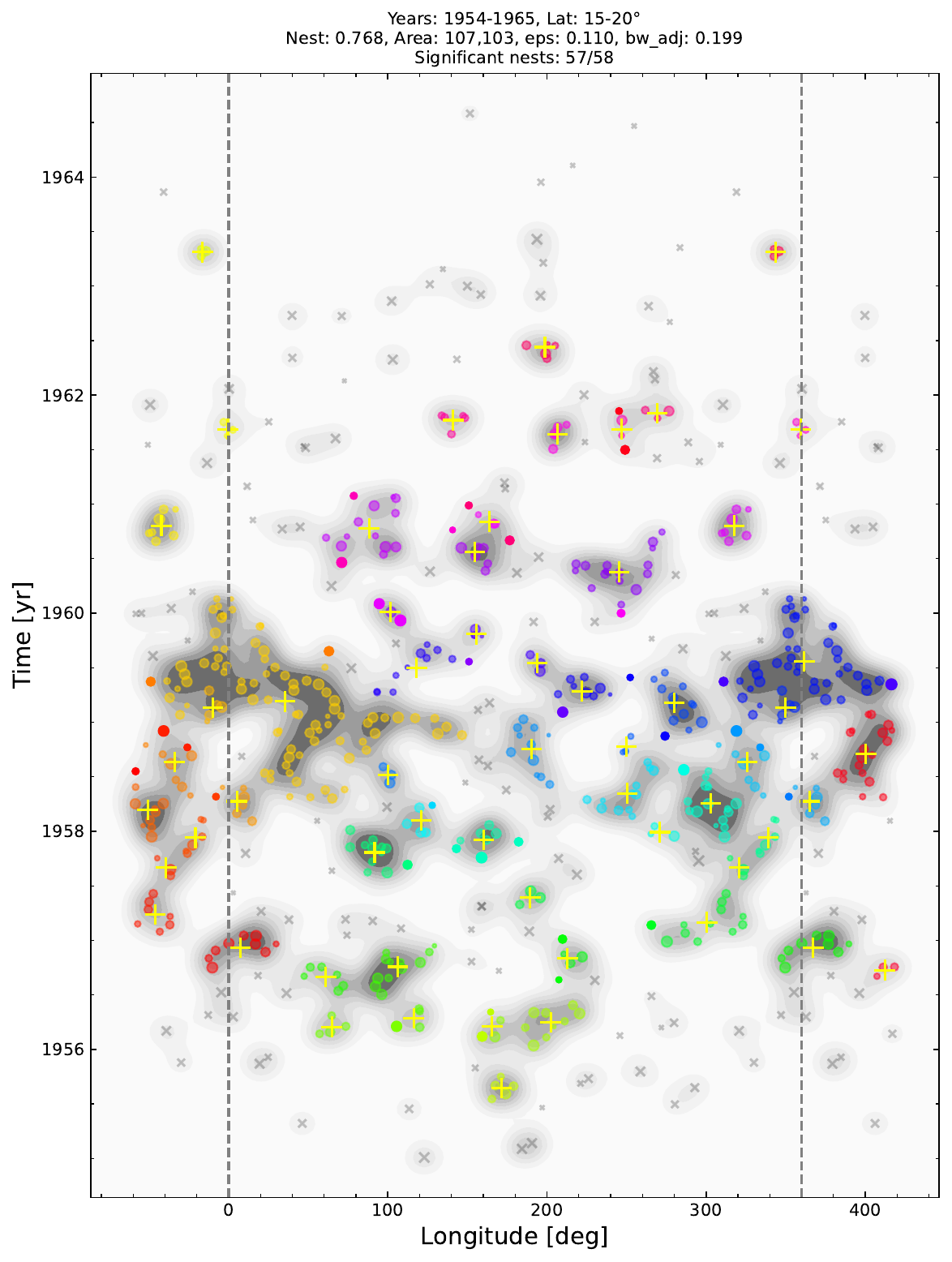}
\includegraphics[width=0.49\textwidth]{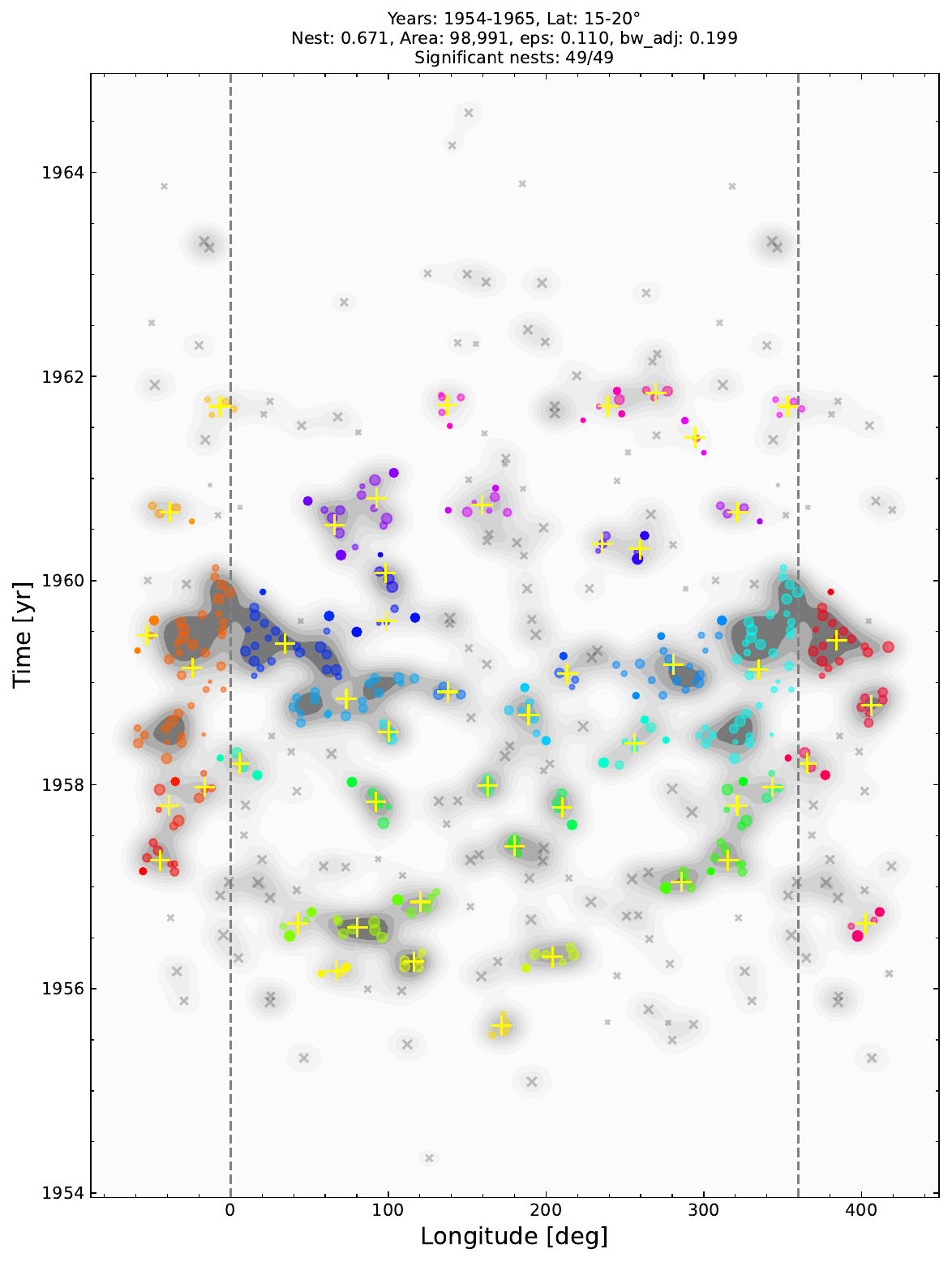}
\caption{Example comparison for Cycle 19, latitude band $15^\circ$--$20^\circ$. 
Left: RGO detects substantially more groups ($D=0.768$). Right: KMAS 
shows sparser coverage ($D=0.671$) for the identical temporal and spatial 
window. Both datasets identify similar large-scale nesting structures, 
but RGO's superior completeness captures additional small groups that 
increase the recovered nesting degree.}
\label{fig:example_comparison}
\end{figure}

This completeness effect remains 
important even at high %when the overall 
nesting degrees, where numerous 
small groups cluster around 
large active regions. Figure~\ref{fig:example_comparison} 
illustrates 
this effect for a representative window (Cycle 19, $15^\circ$--$20^\circ$ 
latitude): RGO identifies substantially more groups, resulting in higher 
recovered nesting degrees in both hemispheres ($D_{\rm RGO} = 0.768$ 
vs. $D_{\rm KMAS} = 0.663$ for the northern band).

The strong correlation validates that our fixed-parameter DBSCAN 
methodology captures reproducible nesting patterns across independent 
observational programmes. For the combined analysis spanning 1874--2025, 
we use RGO data for Cycles 12--20 (1874--1976) and KMAS data for 
Cycles 19--25 (1954--2025). When comparing results across the 
transition, Eq.~(\ref{eq:rgo_kmas_calibration}) provides the 
empirical transformation between the two scales.

\subsection{Spatio-temporal Organisation of Nesting}
\label{ssec:spatial_patterns}

%Having established the long-term 
We have shown that nesting persists
consistently across 14 solar 
cycles.
Next, we examine how nesting patterns vary with heliographic position. 
The combined cycle-averaged statistics for the two datasets is presented in Table~\ref{tab:nesting_by_cycle}. 
Spatial analysis reveals organisation in both systematic and apparently random
configurations, persisting throughout the observational record.

% TABLE 1: NESTING BY CYCLE
% ======================================================================
\begin{table*}
% \centering
\caption{Nesting statistics by solar cycle. RGO data (cycles 12--19) and KMAS data (cycles 18--25) with calibration correction applied (Eq.~\ref{eq:rgo_kmas_calibration}). For overlap cycles (18--19), both datasets are shown. All values represent statistically significant clusters ($p < 0.05$).}
\label{tab:nesting_by_cycle}
\begin{tabular}{lccccccc}
\hline\hline
Cycle & Years & Dataset & $N_\mathrm{win}$ & $\langle N_\mathrm{nest} \rangle$ & $\sigma_N$ & $\langle D \rangle$ & $\sigma_D$ \\
\hline
12 & 1878--1890 & RGO & 6 & 42.2 & 24.6 & 0.552 & 0.128 \\
13 & 1890--1902 & RGO & 7 & 54.6 & 39.7 & 0.539 & 0.151 \\
14 & 1902--1913 & RGO & 6 & 52.5 & 34.8 & 0.560 & 0.095 \\
15 & 1913--1923 & RGO & 7 & 64.4 & 45.1 & 0.593 & 0.178 \\
16 & 1923--1933 & RGO & 7 & 62.6 & 43.5 & 0.610 & 0.081 \\
17 & 1933--1944 & RGO & 7 & 68.3 & 44.3 & 0.587 & 0.158 \\
18 & 1944--1954 & RGO & 7 & 68.4 & 46.3 & 0.590 & 0.181 \\
\cline{2-8}
{19} & {1954--1964} & RGO & 7 & 79.9 & 37.3 & 0.643 & 0.092 \\
 & & KMAS & 7 & 64.9 & 37.6 & 0.684 & 0.070 \\
\cline{2-8}
{20} & {1964--1976} & RGO & 7 & 70.7 & 33.1 & 0.640 & 0.072 \\
 & & KMAS & 7 & 60.6 & 37.3 & 0.661 & 0.087 \\
\cline{2-8}
21 & 1976--1986 & KMAS & 7 & 61.4 & 39.4 & 0.671 & 0.103 \\
22 & 1986--1996 & KMAS & 7 & 52.7 & 33.5 & 0.642 & 0.072 \\
23 & 1996--2008 & KMAS & 7 & 51.3 & 27.7 & 0.652 & 0.091 \\
24 & 2008--2019 & KMAS & 6 & 49.3 & 30.0 & 0.664 & 0.063 \\
25 & 2019--2025 & KMAS & 7 & 45.3 & 33.3 & 0.627 & 0.101 \\
\hline
\end{tabular}

{$N_\mathrm{win}$: number of analysis windows (latitude bands); $\langle N_\mathrm{nest} \rangle$: mean number of significant nests per window; $\langle D \rangle$: mean nesting degree; $\sigma$: standard deviation across windows.}
\end{table*}

% TABLE 2: NESTING BY LATITUDE
% ======================================================================

\begin{table}
% \centering
\caption{Nesting statistics by latitude band. Combined RGO and KMAS datasets with calibration applied. Values represent averages across all solar cycles covered by each dataset.}
\label{tab:nesting_by_latitude}
\begin{tabular}{lccccc}
\hline\hline
Latitude & Dataset & $N_\mathrm{win}$ & $\langle N_\mathrm{nest} \rangle$ & $\langle D \rangle$ & $\sigma_D$ \\
Band ($^\circ$) & & & & & \\
\hline
{0--5} & RGO & 9 & 30.7 & 0.575 & 0.064 \\
 & KMAS & 7 & 22.7 & 0.609 & 0.066 \\
\cline{2-6}
{5--10} & RGO & 9 & 87.9 & 0.649 & 0.067 \\
 & KMAS & 7 & 69.0 & 0.709 & 0.033 \\
\cline{2-6}
{10--15} & RGO & 9 & 107.4 & 0.671 & 0.084 \\
 & KMAS & 7 & 92.4 & 0.728 & 0.033 \\
\cline{2-6}
{15--20} & RGO & 9 & 97.7 & 0.677 & 0.070 \\
 & KMAS & 7 & 93.7 & 0.717 & 0.031 \\
\cline{2-6}
{20--25} & RGO & 9 & 65.9 & 0.628 & 0.053 \\
 & KMAS & 7 & 61.0 & 0.682 & 0.029 \\
\cline{2-6}
{25--30} & RGO & 9 & 29.8 & 0.518 & 0.103 \\
 & KMAS & 7 & 31.1 & 0.628 & 0.039 \\
\cline{2-6}
{30--35} & RGO & 7 & 10.9 & 0.374 & 0.172 \\
 & KMAS & 6 & 9.8 & 0.505 & 0.050 \\
\hline
\end{tabular}

{Combined statistics from RGO (cycles 12--19) and KMAS (cycles 18--25).}
\end{table}

\subsubsection{Latitude Dependence}

Table~\ref{tab:nesting_by_latitude} quantifies the strong latitude dependence 
of the nesting degree observed across both datasets. Peak nesting occurs in the 
$10^\circ$--$15^\circ$ latitude band with $\langle D \rangle = 0.690 \pm 0.074$ 
in the combined RGO-KMAS dataset (after applying the calibration correction from 
Equation~\ref{eq:rgo_kmas_calibration}). The adjacent $15^\circ$--$20^\circ$ 
band shows similarly high values with $\langle D \rangle = 0.688 \pm 0.060$. 
This mid-latitude enhancement spans the region where toroidal flux emergence 
rates are highest 
% and where the subsurface magnetic field structure is most coherent
throughout the solar cycle.

\begin{figure}
    \centering
    \includegraphics[width=0.7\linewidth]{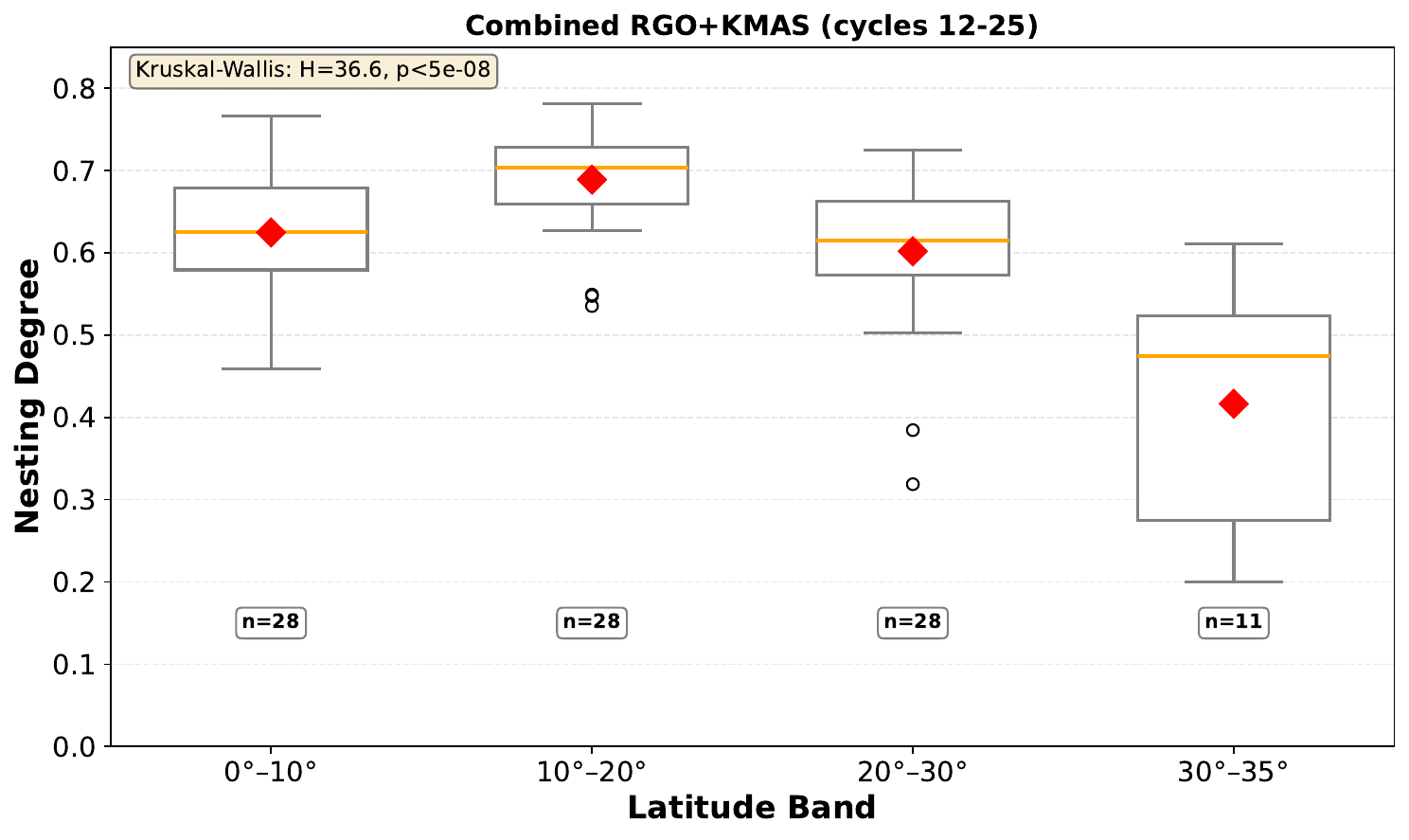}
    \caption{Distribution of nesting degrees across latitude bands from the combined 
RGO (cycles 12--20) and KMAS (cycles 21--25, calibrated) dataset. Box plots show 
distributions across 95 independent full-cycle analysis windows in 10$^\circ$-wide 
latitude bands. Box boundaries mark the first and third quartiles (25th and 75th 
percentiles), with the median indicated by the horizontal orange line and the mean 
by red diamonds. Whiskers extend to 1.5 times the interquartile range, with outliers 
shown as circles. Sample sizes ($n$) indicate the number of analysis windows 
contributing to each band. The Kruskal--Wallis test ($H=36.6$, $p=5.5\times 10^{-8}$) 
confirms highly significant latitude dependence.}
    \label{fig:box}
\end{figure}

Figure~\ref{fig:box} illustrates the distribution of nesting degrees across 
latitude bands using the combined RGO--KMAS dataset. The box plots demonstrate 
remarkably consistent behaviour at mid-latitudes, with the $10^\circ$--$20^\circ$ 
bands showing tight distributions ($\sigma_D \approx 0.07$) that reflect stable 
clustering patterns across all 14 analysed cycles. The median values (horizontal 
lines within boxes) closely track the mean values (red diamonds), indicating 
approximately symmetric distributions at these latitudes. In contrast, the 
$30^\circ$--$35^\circ$ band exhibits substantially larger variance, consistent with the more sporadic emergence 
at high latitudes, particularly near cycle minima.

Nesting degree decreases systematically toward both lower and higher latitudes. 
At the equatorial edge ($0^\circ$--$5^\circ$), the combined dataset yields 
$\langle D \rangle = 0.582 \pm 0.066$, while at high latitudes 
($30^\circ$--$35^\circ$), nesting drops to $\langle D \rangle = 0.416 \pm 0.148$. 
Statistical testing confirms the high significance of this latitude dependence: 
a Kruskal--Wallis test on the combined dataset yields $H = 47.6$ with 
$p = 1.4 \times 10^{-8}$. Furthermore, mid-latitude regions 
($10^\circ$--$20^\circ$) show significantly higher nesting than high-latitude 
regions ($\geq 25^\circ$) with a Mann--Whitney $U$ test giving 
$p = 5.8 \times 10^{-8}$.
\begin{figure}
    \centering
    \includegraphics[width=0.49\linewidth]{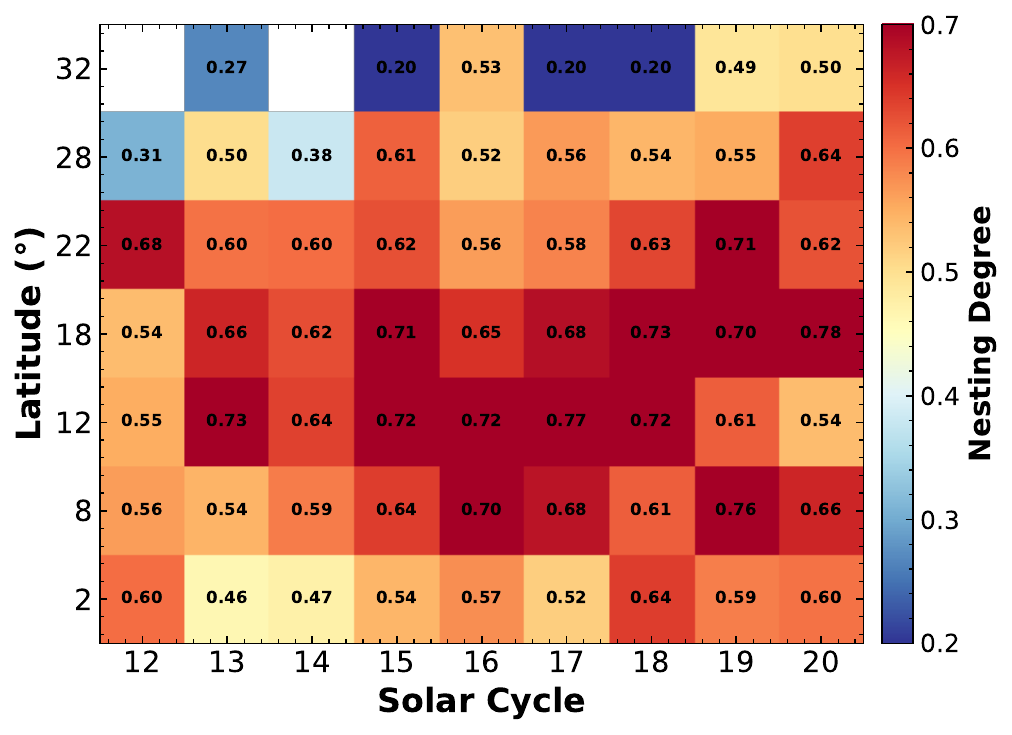}
    \includegraphics[width=0.49\linewidth]{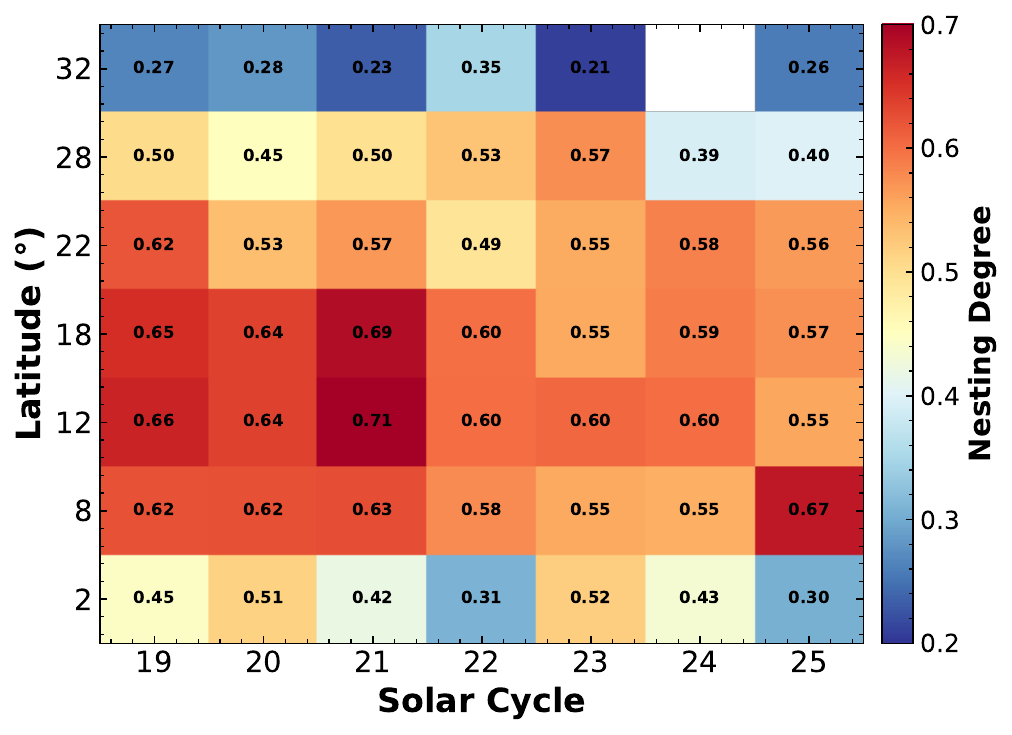}
    \caption{Estimated nesting degree for different 
    cycles in the 5$^\circ$ latitude bands considered, for 
    RGO (left panel) and calibrated KMAS (right panel) datasets. 
    Values are annotated in each cell.}
    \label{fig:heatmap}
\end{figure}

Figure~\ref{fig:heatmap} shows the cycle-to-cycle evolution of this latitude 
pattern for both datasets separately. The persistent mid-latitude enhancement 
appears consistently across all 14 cycles spanning 147 years. Both RGO (left 
panel, cycles 12--20) and KMAS (right panel, cycles 19--25) independently 
reproduce the characteristic pattern, with darker colours (higher $D$ values) 
concentrated in the $10^\circ$--$20^\circ$ bands. Cycle 19 exhibits particularly 
strong nesting ($D > 0.68$) over the $10^\circ$--$25^\circ$ range in both 
datasets, while weaker cycles such as 14 and 24 maintain the mid-latitude peak 
despite lower overall activity levels. The agreement between the two datasets 
after calibration correction demonstrates that the latitude dependence represents 
a fundamental feature of solar magnetic emergence rather than an observational 
artefact.

The physical interpretation of this pattern relates directly to toroidal field structure of the solar 
dynamo. The $10^\circ$--$20^\circ$ bands mark the 
latitude range where:
\begin{enumerate}
\item Toroidal magnetic flux reaches its maximum values during most of 
each cycle.
\item Emergence rates are highest, providing numerous opportunities for 
      repeated flux emergence in clustered patterns.
\item The subsurface magnetic field remains coherent over extended longitudinal 
      ranges, enabling the formation of large-scale active region complexes.
\end{enumerate}

At higher latitudes ($>25^\circ$), both emergence frequency and toroidal field 
coherence decrease, reducing opportunities for nest formation. Near the equator 
($<10^\circ$), while emergence rates remain substantial, the nesting degree 
shows greater cycle-to-cycle variation, possibly reflecting the transition 
between cycles. 
% where emergence from successive dynamo waves can overlap in this latitude range.

\subsubsection{Longitude Distribution}

\begin{figure}
    \centering
    \includegraphics[width=0.49\linewidth]{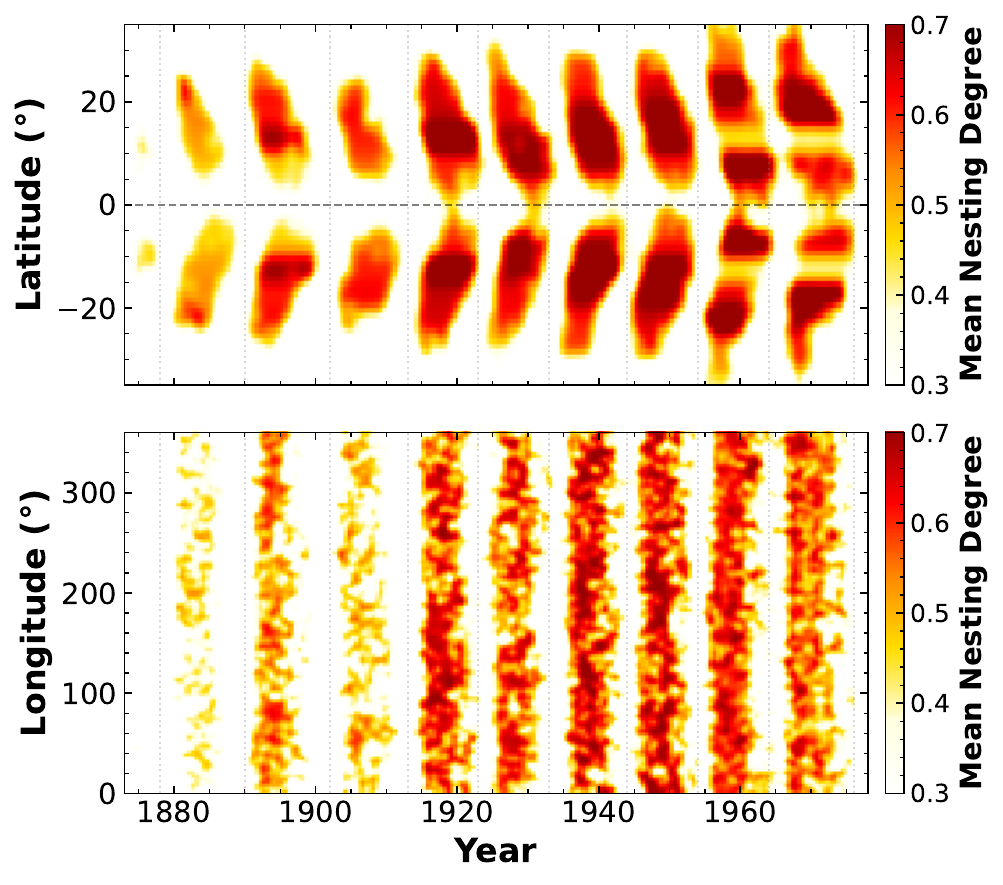}
    \includegraphics[width=0.49\linewidth]{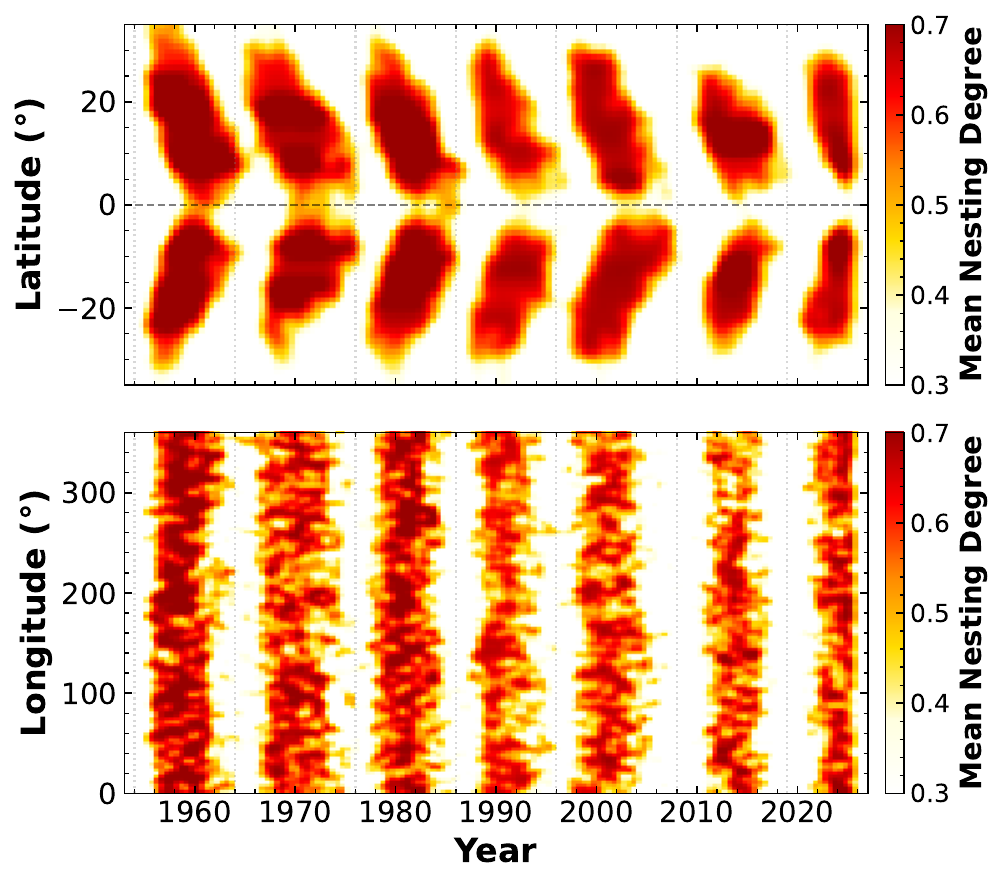}
    \caption{D-weighted KDE showing latitude-time (top) and 
    longitude-time (bottom) projections for RGO (left) and 
    calibrated KMAS (right). }
    \label{fig:butter}
\end{figure}

In contrast to the strong latitude dependence, sunspot group nesting shows 
remarkable uniformity in Carrington longitude. Figure~\ref{fig:butter} shows 
D-weighted kernel density estimates on both the latitude-time (top panels) and 
longitude-time (bottom panels) planes for RGO and KMAS datasets separately.

The latitude-time projections (top panels of Fig.~\ref{fig:butter}) reproduce 
the classic butterfly diagram pattern, with high nesting degrees (red tones, 
$D > 0.65$) concentrated in mid-latitude bands that migrate equatorward through 
each cycle. This visualisation confirms that the latitude dependence documented 
in Table~\ref{tab:nesting_by_latitude} and Fig.~\ref{fig:heatmap} 
persists throughout the temporal 
evolution of each cycle rather than reflecting a time-averaged artefact.

The longitude-time projections (bottom panels) reveal strikingly different 
behaviour. Unlike the strongly structured latitude-time patterns, the 
longitude-time plane shows diffuse, relatively homogeneous nesting patterns 
distributed across all Carrington longitudes. While individual nests exhibit 
clear longitudinal localisation (as shown in Fig.~\ref{fig:nest_examples}), 
the aggregate distribution indicates no preferred emergence longitudes when 
averaged over full solar cycles. This longitudinal uniformity occurs despite 
substantial non-uniformity in the temporal distribution of nesting, with some 
periods showing concentrated emergence and others showing sparse activity.

The homogeneous longitude distribution of nests has important implications:
\begin{enumerate}
\item Nests forming at different 
      latitudes experience different rotation rates in the Carrington frame. 
      Over the 10--12 year duration of each analysis window, this differential 
      rotation effectively smears any initial longitude preferences throughout 
      the full $360^\circ$ range.
      
\item While individual nests can persist 
      for multiple rotations at fixed Carrington longitudes (or exhibit 
      systematic drift patterns), such features appear and disappear on 
      timescales shorter than full solar cycles. The cycle-averaged view thus 
      captures a succession of distinct nesting events rather than persistent 
      active longitude structures.
      
\item Many identified nests exhibit systematic 
      longitude drift independent of the local differential rotation rate 
      (Figure~\ref{fig:nest_examples}). Such drifting nests further contribute 
      to longitudinal homogenisation when integrated over multi-year periods.
\end{enumerate}

This longitude independency distinguishes our full-cycle analysis from shorter 
timescale studies of active longitudes. On rotational timescales 
($\lesssim 1$ year), pronounced longitude preferences are commonly observed 
\citep{Berdyugina03, Usoskin07, Korsos24}. However, these structures evidently lack the 
multi-year persistence necessary to generate cycle-averaged longitude 
asymmetries in nesting patterns.

% Kruskal-Wallis test confirms highly significant latitude 
% dependence ($H = 40.0$, $p < 10^{-6}$).

\subsection{Correlation with Solar Activity}
\label{ssec:activity_correlation}

% correlation of D with total area, cycle amplitude, etc. 

\begin{figure}
    \centering
    \includegraphics[width=0.49\linewidth]{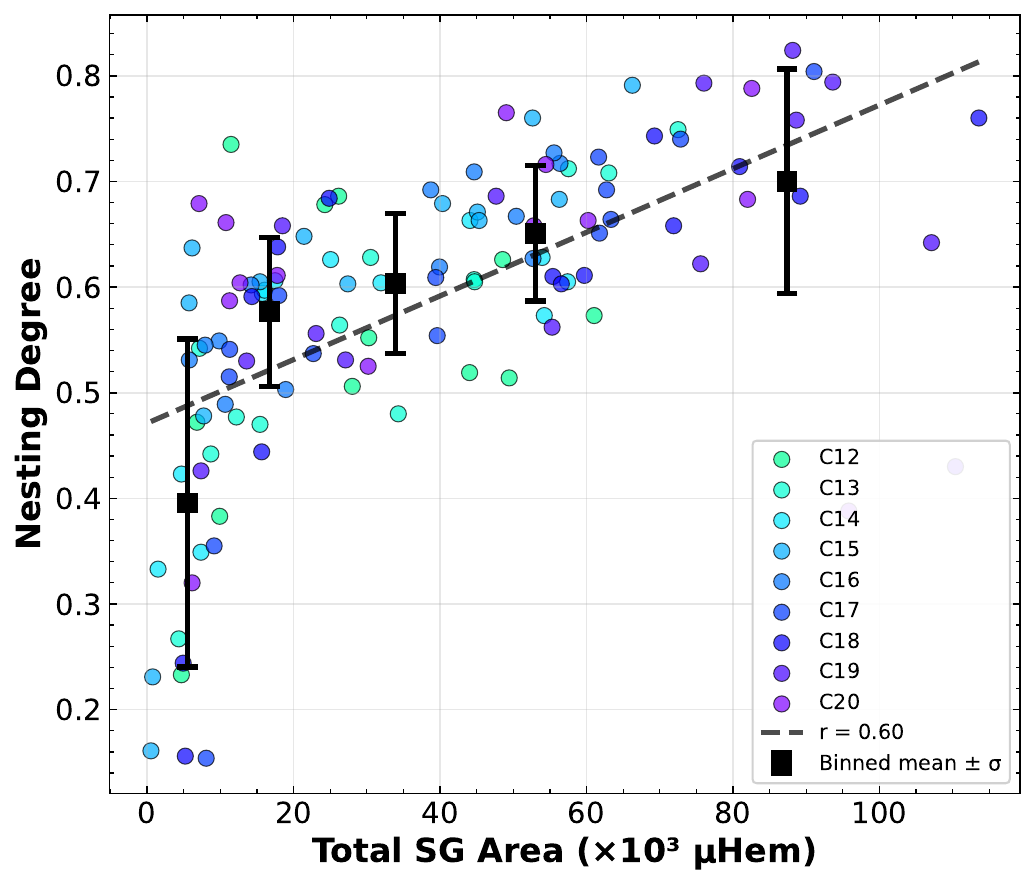}
    \includegraphics[width=0.49\linewidth]{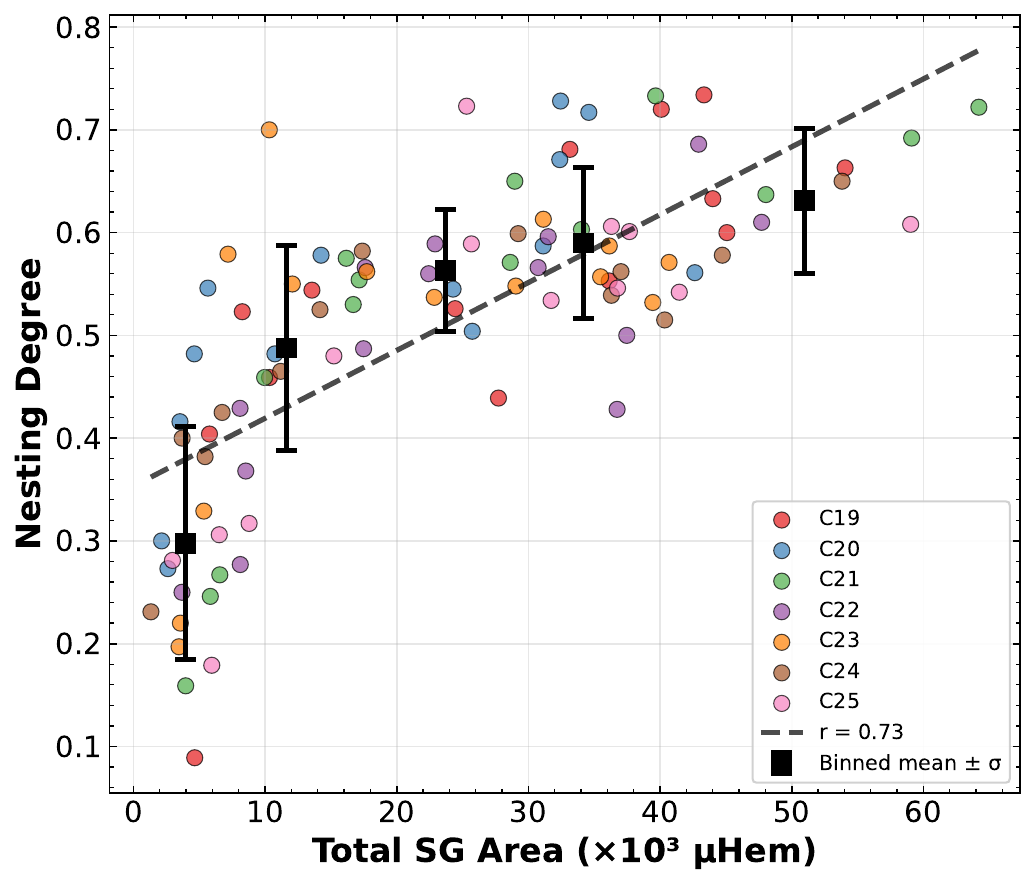}
    \caption{Nesting degree change with solar activity expressed in terms of total SG areas within analysed windows for RGO (left) and KMAS (right).}
    \label{fig:nesting-activity-corr}
\end{figure}

Figure~\ref{fig:nesting-activity-corr} shows the relationship between the
nesting degree and the overall solar activity level, quantified by the total 
SG area within each analysis window. 
We find a clear positive correlation in both datasets, with Pearson coefficients of $r=0.60$ for RGO and $r=0.73$ for KMAS. 
This indicates that stronger cycles, characterised by larger total emergent flux, tend to exhibit higher nesting degrees. 

The trend is not strictly linear and there is considerable cycle-to-cycle scatter, especially in the KMAS dataset where small-number statistics affect low- and high-activity edges.
Nevertheless, the overall behaviour suggests that more coherent toroidal flux systems during strong cycles give rise to more frequent or more pronounced longitudinal clustering.
This finding is consistent with the enhanced large-scale magnetic organisation expected during periods of elevated solar activity. 

For each 
threshold, we recomputed nesting degrees for all 
time-latitude windows and recalculated the correlation with 
activity level.

\subsubsection{Correlation stability}
\label{sssec:corr-stab}
One key question is whether the correlation between nesting degree 
and activity level persists when the analysis is
restricted to larger groups, or whether it arises 
primarily because high-activity periods include more groups 
available to form apparent clusters.
To test this,
%whether the nesting-activity correlation represents 
%genuine physical organisation or a statistical artefact from 
%abundant small groups, 
we repeated the analysis for progressively increasing minimum
area thresholds.

\begin{figure}
    \centering
    \includegraphics[width=\linewidth]{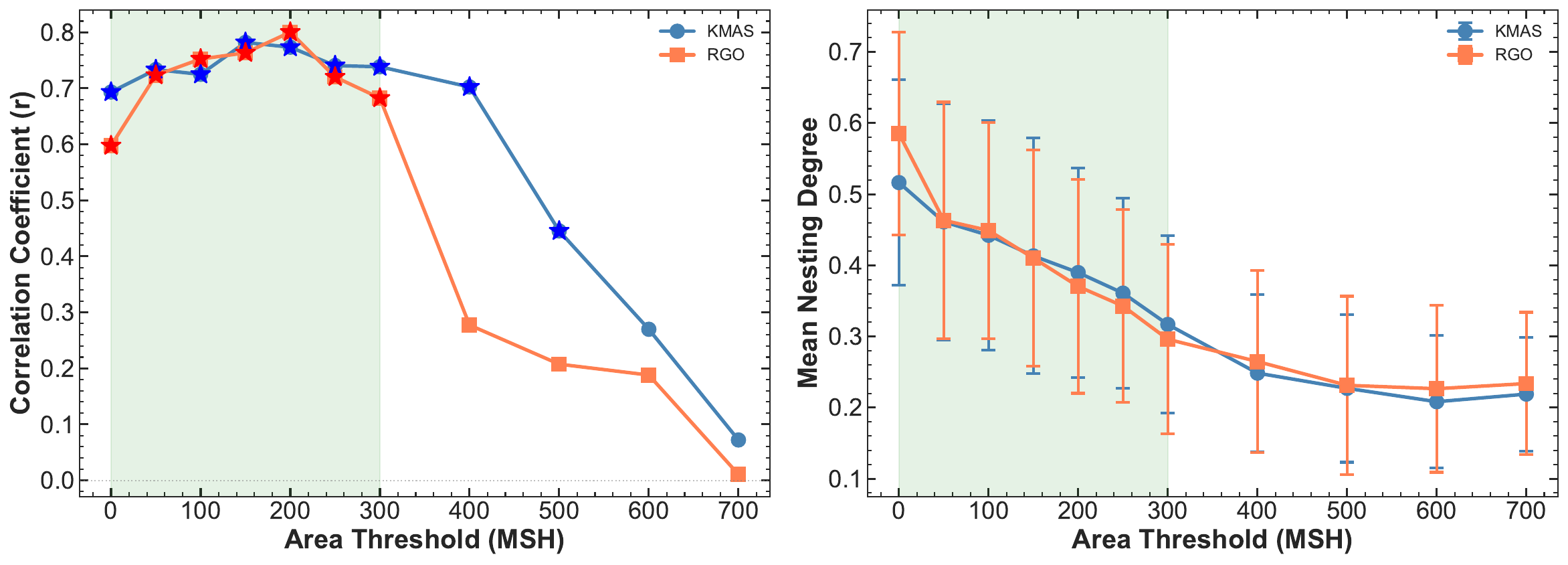}
    \caption{Pearson correlation coefficients between the nesting degree and the 
    activity level (left) and the nesting degree (right) 
    as a function of the minimum area threshold, 
    for KMAS (blue) and RGO (red) datasets. 
    Star symbols denote correlations with $p<0.01$.
    The green region shows the region where the 
    nesting-activity correlation are rather stable. 
    The error bars show $1\sigma$ intervals.}
    \label{fig:corr-stab}
\end{figure}

Figure~\ref{fig:corr-stab} (left panel) shows the correlation 
coefficient between nesting degree and the activity level (total area 
across time-latitude windows) as a function of the area threshold.
For thresholds up to 200 MSH (KMAS) and 300 MSH (RGO), 
both datasets maintain strong, highly significant correlations 
(KMAS: $r = 0.70$-$0.79$; RGO: $r = 0.60$-$0.78$, all $p<10^{-12}$).
Notably, in RGO the correlation \emph{increases} from $r=0.60$ at 
zero threshold to $r=0.78$ at about 200~MSH, indicating that the 
nesting--activity relationship is strongest for medium-to-large groups.
This behaviour argues
%increasing correlation as small groups are excluded provides 
%additional evidence 
against a statistical counting artefact: if the correlation were driven mainly by
abundant 
small groups during high-activity periods, excluding
them would \emph{weaken} the correlation.
Instead, the correlation strengthens, suggesting that the
organisation of medium-to-large active regions 
($\sim$100-300~MSH) plays a dominant role.
At thresholds above 300 MSH, the correlation decreases as 
sample sizes become small.

% \subsubsection{Size-Dependent Nesting Structure}
% \subsubsection{Hierarchical Nesting Structure}
\subsubsection{Nesting Hierarchy}
\label{sssec:hierarchy}

Beyond validating the nesting-activity correlation, the threshold 
analysis in Sect.~\ref{sssec:corr-stab} reveals hierarchical 
organisation in SG emergence. Figure~\ref{fig:corr-stab} 
(right panel) shows mean nesting degree as a function of the area 
threshold for both datasets.
Both KMAS (1955--2025, Cycles 19--25) and RGO (1874--1976, 
Cycles 11--20) exhibit consistent patterns: mean nesting degree 
decreases from $\sim$0.50--0.60 for all groups to $\sim$0.23--0.27 
for groups exceeding 400--450~MSH. 
This decline indicates that 
small SGs are preferentially associated with clustered emergence,
whereas large regions alone exhibit weaker clustering.

The non-zero asymptotic nesting degree of $\sim$0.25 at high 
thresholds indicates that even large active regions retain some tendency 
to form nests. The substantially higher nesting 
degree at low thresholds ($\sim$0.50) reflects `parasitic' flux 
emergence, where small groups preferentially emerge near existing large 
active regions, possibly led by fragmentation of large-scale toroidal 
flux loops into smaller parts, clustering around the nest centre. 

Overall, these results are consistent with a hierarchical emergence 
pattern: (1) small, parasitic groups ($<$200~MSH) preferentially emerge near 
large active regions, producing strong clustering, and 
(2) large active regions themselves ($>$300~MSH) emerge with reduced but persistent longitudinal 
organisation \citep[see also][Sect. 3.3]{Jiang11}. 

The error bars in Fig.~\ref{fig:corr-stab} represent 1$\sigma$ 
standard deviation across time-latitude windows, reflecting variation 
with solar activity level and cycle phase. The large error bars 
($\sim$0.10--0.20) predominantly in the low-threshold regime, 
demonstrate substantial window-to-window variation, which is the 
source of the nesting-activity correlation presented in
Sect.~\ref{sssec:corr-stab}. 

\subsection{Inter-nest Length Scale}
\label{ssec:internest}

To characterise the spatial organisation of nesting, we compute an inter-nest 
length scale for each analysis window ($5^\circ$ latitude band over a full 
solar cycle), 
\begin{equation}
    \lambda_{\rm typ} = (A / N_{\rm sig})^{1/2},
\end{equation}
where $A=360^\circ\times 5^\circ$ is the angular area of the latitude 
band, and $N_{\rm sig}$ is the number of statistically significant nests 
identified in that window. This metric represents the mean angular separation 
obtained by uniformly distributing $N_{\rm sig}$ nests across the available 
longitude range. We convert this to a physical length scale 
$L_{\rm typ}$ (Mm) using the central latitude of each band.

\begin{figure}
    \centering
    \includegraphics[width=0.7\linewidth]{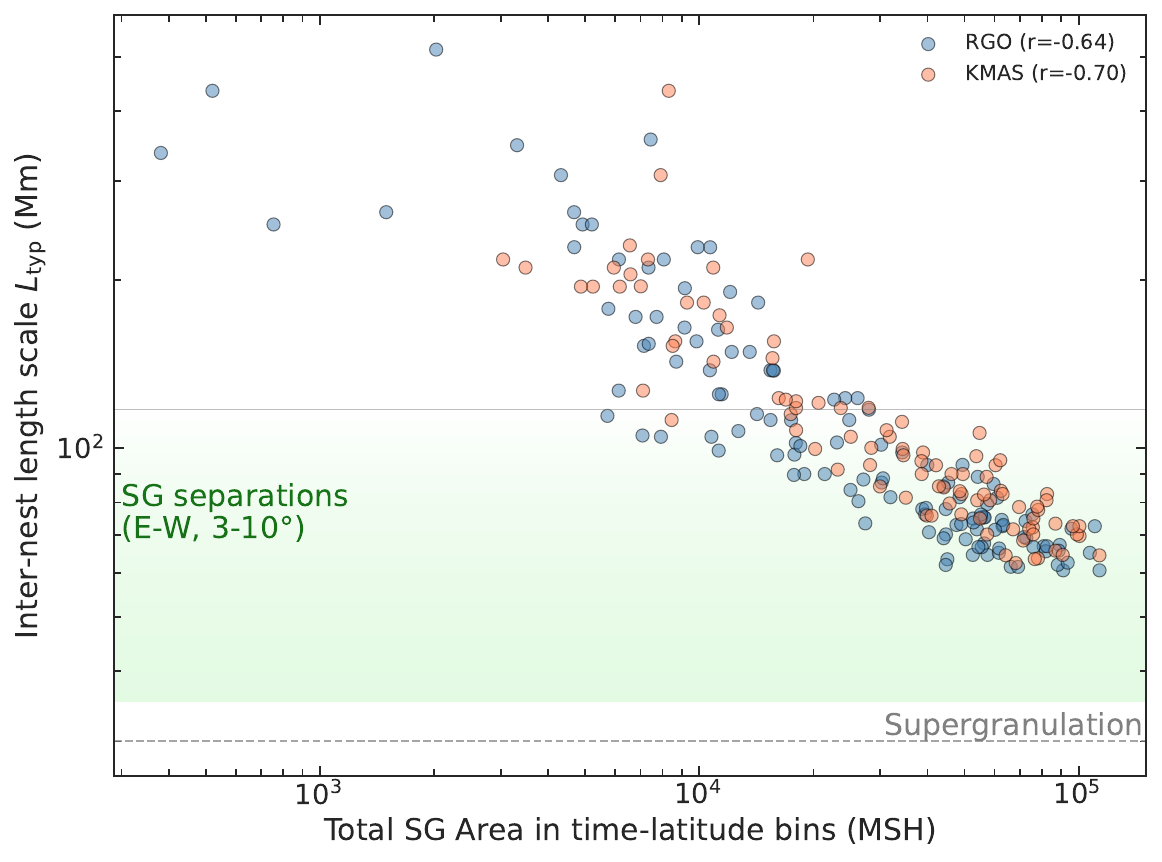}
    \caption{Inter-nest length scale $L_{\rm typ}$ as a function 
    of the overall activity level in terms of the total SG area in each 
    window considered. The range of SG longitudinal (E-W) separations is 
    marked with a green band, where the gradient qualitatively indicates 
    the size-dependent emergence likelihood.}
    \label{fig:lengthscale}
\end{figure}

Figure~\ref{fig:lengthscale} shows the inter-nest length scale as a function 
of total SG area (a proxy for solar activity level) for both RGO (1874--1976) 
and KMAS (1954--2025) latitude-time windows. Both datasets exhibit a strong 
negative correlation between $L_{\rm typ}$ and activity level, with Pearson 
coefficients $-0.641$ ($p\sim 10^{-16}$) for RGO and $-0.700$ 
($p\sim 10^{-15}$) for KMAS. At low activity levels (sparse emergence), 
$L_{\rm typ}$ reaches 200--500~Mm, while at high activity levels it 
converges to $\sim$60--100~Mm, comparable to typical SG elongations in 
the E-W direction (3--10$^\circ$, corresponding to $\sim$35--117~Mm at 
$15^\circ$ latitude). The mean value of $L_{\rm typ}$ is approximately 
120~Mm for both datasets.

However, $L_{\rm typ}$ represents a phase--space density measure that 
integrates over the full solar cycle duration, including periods of minimal 
or no emergence. Since nests preferentially cluster in active longitude 
zones during emergence \citep[e.g.,][]{gai83}, the actual local spacing 
between nests within these zones is likely substantially smaller than 
$L_{\rm typ}$. Nevertheless, the strong anti-correlation with 
activity level demonstrates that nests pack more compactly in phase space 
during solar maximum, with the convergence to SG elongation scales 
suggesting dense spatial organisation during peak activity.

\section{Discussion}
\label{sec:discuss}

Our analysis provides the first fully automated, cycle-by-cycle quantification of sunspot-group nesting using density-based clustering on a century-scale dataset.
Several qualitative results long recognised in earlier studies naturally emerge from our DBSCAN-based approach.

First, the typical scale and temporal persistence of nests identified here align well with the classical findings of \citet{cas86} and \citet{bro90}.
Nests commonly persist for one to several solar rotations and can extend across tens of degrees in Carrington longitude, consistent with the recurrent emergence of magnetic flux from long-lived subsurface concentrations.
The observed displacement of nest centres in longitude, often independent of the local differential rotation, matches earlier descriptions of active-region complexes and reflects underlying magnetic structures that rotate with their own characteristic phase speeds.

Second, the latitude dependence of the nesting degree provides an important constraint on models of flux emergence.
We find the highest nesting degree ($D\approx 0.6$) within $10^\circ -20^\circ$ latitude, where the majority of active regions appear during the maximum and declining phases of cycles.
Lower values toward higher latitudes primarily reflect reduced emergence rates rather than a weakening of the clustering tendency itself.
This behaviour mirrors the butterfly diagram and the decreasing coherence of toroidal flux systems near cycle onset.

Recent studies using density-based clustering identified persistent active 
longitudes ($\sim 40^\circ$ wide, 15-25 Carrington rotations) by analysing all latitudes 
together \citep{Korsos24,Csaszar25}. Our latitude-sliced 
approach quantifies nesting in both longitude and latitude over full solar cycles, 
revealing strong mid-latitude enhancement but no persistent longitudinal 
preference (Fig.~\ref{fig:butter}). This cycle-scale longitudinal homogeneity reflects 
differential rotation dispersing shorter-lived active longitude structures 
across all longitudes over decadal periods.

The third result comes from analysing the RGO and KMAS catalogues independently over their overlap period. 
RGO consistently yields slightly higher nesting degrees, which we attribute to its more complete sampling of small sunspot groups. 
Despite this offset, both datasets display the same latitude dependence, similar ranges of $D$, and comparable cycle-to-cycle variability. 
This agreement indicates that the clustering patterns recovered by our method represent intrinsic properties of solar emergence rather than artefacts of a particular catalogue.

Fourth, the synthetic-data validation demonstrates that the chosen clustering parameters recover the true nesting degree with $\lesssim 10$\% scatter.
Importantly, the behaviour of DBSCAN differs from classical active-longitude analyses, which typically rely on sinusoidal fits or phase-folding of emergence patterns.
DBSCAN captures clustering of arbitrary shape, enabling us to detect both compact nests and extended, drifting structures that would be difficult to characterise using traditional methods.

A further result is the moderate but clear positive correlation between nesting degree and overall cycle activity (Section~\ref{ssec:activity_correlation}). 
Stronger cycles, characterised by larger emergent flux, tend to show higher clustering. 
This behaviour is consistent with the idea that more coherent toroidal flux systems during strong cycles feed repeated localised flux emergence, whereas weaker cycles exhibit reduced magnetic organisation. It is also consistent with a similar possibility proposed for solar-type stars, where an increase in the nesting degree with the activity level could explain the observed trends in photometric variability amplitudes \citep{isik20}. 

Finally, these results have implications for Sun-as-a-star variability studies. 
If more than half of all sunspot groups emerge within nests, as our analysis indicates, then the rotational modulation of solar irradiance is strongly influenced by nest persistence and longitudinal asymmetry.
This connects directly to stellar observations, where active longitudes and long-lived spot complexes are commonly inferred in photometric time series \citep{Ozavci18,Breton24}.
The quantitative nesting degree ($D \approx 0.61$) provides a benchmark for 
comparison with stellar active-region nesting \citep{isik20}. If similar nesting fractions 
occur on other Sun-like stars, photometric variability and rotational 
modulation amplitudes should reflect both the instantaneous distribution of 
active regions and their tendency to cluster in spatial-temporal complexes.

A number of general limitations should be noted.
DBSCAN relies on a single spatial–temporal scale parameter, which cannot capture variations in nest size within an analysis window.
Our approach also treats longitude and time symmetrically after standardisation, whereas the physical processes governing flux emergence may introduce anisotropy. Furthermore, the longitude–time projection does not account for potential clustering in latitude.
Future work could incorporate full 3D clustering or probabilistic models that allow continuous tracking of nest evolution.

The inter-nest length scale $L_{\rm typ}$ (Sect.~\ref{ssec:internest}) represents a phase-space density measure 
integrating over full cycle durations and uniform spatial distribution across 
$360^\circ\times5^\circ$ bands. Consequently, $L_{\rm typ}$ does not reflect 
actual local spacing between neighbouring nests within active zones, where emergence 
concentrates preferentially \citep{gai83}. Nevertheless, the robust 
anti-correlation with activity (Fig.~\ref{fig:lengthscale}) reveals 
a significant pattern: during solar maximum, nests pack into denser configurations, approaching typical SG elongations. This convergence may reflect hierarchical magnetic organisation, 
with compact nests clustering into larger composite structures spanning up to 
$55^\circ$ longitude \citep{bro90}. Refined analysis focusing on active 
emergence periods rather than cycle-integrated densities would better characterise 
local inter-nest spacing but lies beyond the scope of the present study.

Taken together, our results show that nests constitute real, temporally coherent structures on timescales of several rotations, yet their collective contributions do not produce long-lived active longitudes when averaged over full solar cycles. 
Differential rotation and intrinsic longitudinal drift disperse nests on timescales shorter than a cycle, preventing persistent non-axisymmetric patterns.

\section{Conclusions}
\label{sec:conclu}

We presented an automated, data-driven approach to identify sunspot-group nests in the longitude--time domain and applied it independently to the RGO (1874--1976) and KMAS (1955--2025) catalogues.
The method reliably detects coherent emergence patterns and is supported by tests on synthetic data.

Our main findings are as follows. 
\begin{itemize}
    \item[--]
Nests are a persistent feature of solar activity:
across all cycles and latitude bands, the mean nesting degree is  $\langle D\rangle =0.613\pm 0.118$, indicating that roughly 61\% of all sunspot groups emerge within clusters. \\
    \item[--]
Nesting exhibits a clear latitude dependence, peaking ($D\approx 0.69$) between 10$^\circ$ and 15$^\circ$, consistent with densest phase of the butterfly diagram and the organisation of toroidal flux in the activity belts. 
 \\
    \item[--]
    Nesting degree correlates positively with overall cycle strength, suggesting that more coherent toroidal flux bundles sustain repeated flux emergence. Importantly, this correlation strengthens when small groups are progressively excluded, indicating that nesting represents genuine clustering in flux emergence rather than statistical artefacts from abundant small groups.
 \\
    \item[--] Although individual nests can remain coherent for several rotations, differential rotation and longitudinal drift disperse them on timescales short compared to a solar cycle, preventing long-lived active longitudes in cycle-averaged emergence. 
 \\
\item[--]
Nesting exhibits hierarchical structure: small sunspot groups cluster strongly 
around large active regions (nesting degree $\sim$0.5--0.6), while large groups 
emerge with greater spatial independence (asymptotic nesting degree $\sim$0.25).
\\
\item[--]
Characteristic inter-nest spacing shows strong anti-correlation with activity, contracting from $\sim$200--500~Mm during low activity 
to $\sim$60--100~Mm during solar maximum, comparable to typical sunspot group 
elongations (35--117~Mm) and indicating dense spatial packing of emergence 
sites.
\\
    \item[--] Despite quantitative differences, RGO and KMAS yield qualitatively consistent nesting picture, suggesting the robustness of our approach. Quantitative differences are at least partially due to non-exhaustive observational coverage. 
\end{itemize}

Our findings demonstrate that unsupervised machine-learning–based clustering offers a powerful unified approach for quantifying nesting and for linking solar flux-emergence patterns to long-term magnetic activity.
Extensions of this approach to full 3D clustering and to stellar photometric data may provide new constraints on the magnetic activity of Sun-like stars.

\begin{acks}
 We thank the anonymous reviewer for their comments and suggestions. 
\end{acks}

%%% %%%%%%%%%%%%%%%%%%%%%%%%%%%%%%%%%%%%%%%%%%%%%%%%%%%%%%%%%%%
%% Bibliography
%
% Using BibTeX
%
\bibliographystyle{spr-mp-sola}
\bibliography{main.bib}

\end{document}